# Timing is everything: How subtle timing changes in MRI echo planar imaging can significantly alter mechanical vibrations and sound level

Amir Seginer[1,2], Alexander Bratch[3], Shahar Goren[2,4], Edna Furman-Haran[1,2], Noam Harel[3], Essa Yacoub[3], Rita Schmidt[2,4]

[1]Life Sciences Core Facilities, Weizmann Institute of Science, Israel

[2]The Azrieli National Institute for Human Brain Imaging and Research, Weizmann Institute of Science

[3]Center for Magnetic Resonance Research (CMRR), University of Minnesota, Minneapolis, Minnesota USA

[4]Department of Brain Sciences, Weizmann Institute of Science, Israel

**Abstract**

Modern MRI relies on the well-established Echo-Planar-Imaging (EPI) method for fast acquisition. EPI is the workhorse of diffusion and functional MRI in neuroscience as well as of many dynamic applications for clinical body imaging. Its speed stems from rapidly switching currents through the 'gradient coils' responsible for spatial encoding. These quick changes, within a strong static magnetic field induce Lorentz forces that generate mechanical vibrations, leading to the loud, characteristic MRI noise. This acoustic noise is already a significant concern in standard clinical scanners, requiring hearing protection and risking image degradation or even hardware strain. In ultra-high-field systems (≥7T), the issue is exacerbated due to stronger Lorentz forces. In this study, we introduce a novel model that characterizes the acoustic spectrum for a given EPI scan. The spectrum results from interference between multiple identical periods of alternating currents, making the relative timing between them a key factor. We show that even subtle timing changes significantly alter the sound level. Scans of either high spatial resolution or high temporal resolution, on clinical 7T and investigational 10.5T human scanners, corroborated the model and its predicted acoustics spectra. Acoustic energy changes of up to 47-fold were reached in close to mechanical resonances and up to 5-fold in other regions. Intriguingly, under certain conditions, doubling the acquisitions per unit time actually reduced the minimal acoustic energy two-fold. Not less important, 'ghosting-artifacts' exhibit strong dependence on the scan's acoustic characteristics and the benefit of subtle time delays of internal correction-acquisitions ('navigators') is also demonstrated.

# Introduction

Modern MRI uses widely the well-established Echo-Planar-Imaging (EPI) pulse sequence and its variants for rapid image acquisition. EPI is the workhorse of diffusion and functional MRI (fMRI) widely used in neuroscience[1], and of many dynamic applications for clinical body imaging[2,3]. The power of EPI[4] is its ability to quickly acquire the whole 3D volume of interest, typically within a second or two. To achieve this, the magnetic field 'gradient' coils, which encode position, are quickly switched, alternating the current through them within 0.5–2 ms, typically. This alternating current flows within the strong static magnetic field of the MRI, denoted as $B_0$, which applies a Lorentz force on the current. Since this is an alternating current, the forces on the conductors also alternate, leading to mechanical vibrations of the system. These vibrations produce the distinct sound of the MRI scans, which is especially intense in EPI. The high intensity of this sound requires subjects to wear protective gear (earphones and/or ear plugs). In addition, the vibrations intensity can potentially damage the system. Specifically, the damage can occur when the alternating current frequency is close to a mechanical resonance of the system. This issue is further exacerbated on ultra-high-field scanners (with a static magnetic field $\geq$ 7T), as the much stronger magnetic field leads to much stronger Lorentz forces.

Beyond being a source of significant acoustic noise, EPI sequences are also inherently prone to "ghosting" artifacts in the reconstructed images. These artifacts commonly arise from subtle inconsistencies between the positive and negative gradient lobes used during spatial encoding. To mitigate these inconsistencies, 'navigator' acquisitions are typically employed prior to the main scan, often involving three alternating gradients. Existing evidence suggests a direct correlation between the loud operational characteristics of EPI and the severity of these ghosting artifacts[5].

Hardware solutions to reduce mechanical vibrations of a gradient system and the resulting sound are extensively studied[6–9]. A recent work[7] showed that replacing the solid copper shield of the RF coil with a mesh material reduced the sound levels. At ≥7T systems, third-order shim coil configurations showed a major impact[10], with disconnecting of the third-order shim suppressing part of the resonant peaks.

In software, the current approach to protect the system from reaching mechanical resonances is to simply block certain (gradient) current-alternation rates. However, the dependency between the acoustic spectra and the chosen current-alternation rate of the scan is not straightforward. Examining the acoustic spectra of EPI scans reveals multiple peaks, without the current-alternation frequency necessarily being a dominant one. This is because in modern EPI scans the trains are shorter than in the past, due to contemporary high-acceleration techniques[11]. As a consequence, the interaction between multiple such trains during the scan creates an interference which drives the actual acoustic spectrum. The actual acoustic spectrum is also affected by the mechanical resonances which amplify certain frequencies.

As this acoustic spectrum is due to interference between multiple short trains, the relative timing between them can strongly affect the actual spectrum. Furthermore, as shown here, even subtle timing changes can make a significant difference. The effect of timing in EPI have been considered in the past[12–14], but in those cases the train itself was modified, either changing the envelope of the train and consequently prolonging it three-fold[12,13], or alternating the current in irregular steps[14]. In both cases these modifications make the reconstruction much harder as slight imperfections affect the quality of the final image. Other studies also examined changing the gradient waveform. This includes sine and spiral waveforms which can produce quieter scans[15–18], however, this also results in longer acquisitions. Other works showed iterative solutions solving an inverse problem and aiming to reduce the sound and the gradient slew rate[19]. While these works offer valuable solutions, they usually require special reconstruction approaches.

In this study, we introduce an analytical model to understand and characterize the resulting spectra in multi-train EPI-like acquisitions. We show that modifying the timing *between the trains* enables a degree of freedom that does not encumber the image reconstruction or the choice of acquisition parameters, while providing means to control the acoustic energy. Rigorous comparisons of the model and the measurement were performed. Our model paves the way towards better methods of reducing the vibrations and the sound in EPI scans. In addition, we explore the correlation of the ghost-artifacts intensity with the scan's acoustic characteristics, demonstrating how slight timing changes can reduce the artifacts.

While the new model is applicable across all magnetic field strengths, emerging studies that push the boundaries of fast and high-resolution imaging[20–23] stand to benefit most from this approach, as they are constrained by either acoustic limitations or image quality. These include fast whole-brain acquisitions[20,24,25] important for cognitive neuroscience, multi-echo EPI[26,27] used to characterize physiological and neurovascular responses, and short echo times[21,23,28] that enhance signal-to-noise ratio (SNR). Accordingly, this study included scans on both a clinical 7T and an investigational 10.5T human MRI scanner, validating the model and examining similarities and differences between the setups.

## Acoustic Energy Prediction

### The Analytic Model

An MRI system includes three gradient coils, each generating a magnetic field whose intensity is linear along the different axes $\hat{x}$, $\hat{y}$, and $\hat{z}$, where $\hat{z}$ traditionally points along the uniform and static main magnetic field, denoted as $B_0$. Thus the magnetic field parallel to $\hat{z}$, $B_z(\vec{r},t)$, at any point in space and time is assumed to obey

$$B_z(\vec{r},t) = B_0 + G_x(t)x + G_y(t)y + G_z(t)z = B_0 + \vec{G}(t) \cdot \vec{r}. \qquad (Eq.1)$$

The linear combination $\vec{G}(t)$, simply referred to as a 'gradient', can apply a magnetic field gradient along any direction. Up to minor corrections, the gradients $G_{x/y/z}(t)$ are proportional to the current flowing through them. Therefore, from here on only the gradients are considered, not the currents.

As mentioned above, EPI features a train of alternating gradients. During this train the gradient repeatedly deviates from zero and returns to zero but alternates in sign each time. The time between the center of one such gradient 'lobe' to the next, of opposite sign, is called an echo spacing (ESP). The term 'echo' is due to a local maximum in the received signal at the center of these lobes, where a 'gradient echo' occurs.

To estimate the acoustic spectrum of a multi-slice 2D EPI scan, a simplified model of the scan is used here, disregarding all gradients applied during the scan, except for the alternating gradient train. The time distributions of these gradients trains $G(t)$ is constructed starting from an infinite train $G^{(\infty)}(t)$, cut down to a single finite echo train length (ETL) of duration $T_{\text{ETL}}$ — by multiplying $G^{(\infty)}(t)$ by a boxcar function $\Pi(t/T_{\text{ETL}})$. Optionally, if multiple echo-times (TEs) are acquired per slice, each train is repeated $N_{\text{TE}}$ times every $\Delta\text{TE}$ interval. Note that the TEs should not be confused with the echoes mentioned previously, occurring at the center of each gradient lobe. The echoes per gradient *lobe* are *local* maxima of signal, while the multiple TEs are the times of the *largest* local maxima over each gradient train. The $N_{\text{TE}}$ trains are repeated themselves $N_{\text{slice}}$ times, every $\Delta T_{\text{slice}}$ interval, for each acquired slice in the volume scanned. Finally, in dynamic acquisition such as in fMRI, all slices are repeatedly acquired $N_{\text{TR}}$ times, every repetition time (TR), to collect the time evolution of the dynamic information. Analytically, the above construction of the multiple gradient-trains $G(t)$ can be written as

$$G(t) = \left[G^{(\infty)}(t) \cdot \Pi(t/T_{\text{ETL}})\right] * \left[\sum_{n_{\text{TE}}=0}^{N_{\text{TE}}-1} \delta\left(t - T_{0,\,\text{TE}} - n_{\text{TE}} \cdot \Delta TE\right)\right] *$$
$$\left[\sum_{n_{\text{slice}}=0}^{N_{\text{slice}}-1} \delta\left(t - T_{0,\,\text{slice}} - n_{\text{slice}} \cdot \Delta T_{\text{slice}}\right)\right] * \quad\quad\quad \text{(Eq.2)}$$
$$\left[\sum_{n_{\text{TR}}=0}^{N_{TR}} \delta\left(t - n_{\text{TR}} \cdot TR\right)\right],$$

where $*$ denotes a convolution, $\Pi(t)$ is the boxcar function, defined here as

$$\Pi(t) \equiv \begin{cases} 1 & 0 < t < 1 \\ 0 & \text{otherwise} \end{cases},$$

$\delta(t)$ is the Dirac delta function, and $T_{0,TE}$ and $T_{0,\,\text{slice}}$ are fixed temporal offsets.

It is implicitly assumed that

$$T_{\text{ETL}} \cdot \frac{1}{\pi} \omega_{2\text{ESP}} = \text{integer},$$

where

$$\omega_{2\text{ESP}} \equiv \frac{2\pi}{2\text{ESP}},$$

or simply, that the echo-train consists of an integer number of gradient lobes (pairs of positive and negative lobes and possibly an extra positive/negative one).

Performing a Fourier transform and using the convolution theorem (see SI Appendix S6), the angular frequency decomposition $g(\omega)$ of the gradient can be shown to be

$$g(\omega) = \left\{g^{(\infty)}(\omega) * \left[T_{\text{ETL}} \cdot e^{-i\omega T_{\text{ETL}}/2} \text{sinc}\left(\tfrac{1}{2}\omega \cdot T_{\text{ETL}}\right)\right]\right\} \times$$
$$e^{-i\omega \cdot T_{0,TE}} e^{-i\omega \cdot \Delta TE \cdot (N_{\text{TE}}-1)/2} \frac{\sin(\omega \cdot \Delta TE \cdot N_{\text{TE}}/2)}{\sin(\omega \cdot \Delta TE/2)} \times$$
$$e^{-i\omega \cdot T_{0,\text{slice}}} e^{-i\omega \cdot \Delta T_{\text{slice}} \cdot (N_{\text{slice}}-1)/2} \frac{\sin(\omega \cdot \Delta T_{\text{slice}} \cdot N_{\text{slice}}/2)}{\sin(\omega \cdot \Delta T_{\text{slice}}/2)} \times \quad \text{(Eq.3)}$$
$$e^{-i\omega \cdot TR \cdot (N_{\text{TR}}-1)/2} \frac{\sin(\omega \cdot TR N_{\text{TR}}/2)}{\sin(\omega \cdot TR/2)},$$

where the sine ratio of the form

$$\frac{\sin(\alpha N)}{\sin(\alpha)}$$

is sinc-like in $\alpha$, but cyclic, with a peak of height $N$ and width $\Delta\alpha = 2\pi/N$, repeating (up to a sign) every $\alpha = \pi n$ ($n = 0, \pm 1, \pm 2, \ldots$).

One simple infinite gradient train $G^{(\infty)}$ is a sinusoidal gradient

$$G^{(\infty)}_{\text{sine}}(t) \equiv G\sin(\omega_{2\text{ESP}} t),$$

however, modern scans usually use symmetric trapezoid gradients. A single symmetric trapezoid can be defined as a convolution of two boxcars, of widths $\tau_{\text{ramp}}$ (the time from gradient zero to the maximal gradient, in absolute value) and $\tau_{\text{ramp}} + \tau_{\text{flattop}}$, where $\tau_{\text{flattop}}$ is the duration during which the gradient is maximal, before dropping off again. Thus, an infinite train of alternating trapezoid gradients can be written as

$$G_{\text{trap.}}^{(\infty)}(t) = \left[\frac{G}{\tau_{\text{ramp}}} \Pi(t/\tau_{\text{ramp}}) * \Pi\left(t/(\tau_{\text{flattop}} + \tau_{\text{ramp}})\right)\right] * \quad \text{(Eq.4)}$$
$$\left[\sum_{n=-\infty}^{\infty} \delta(t - 2n \cdot \text{ESP}) - \sum_{n=-\infty}^{\infty} \delta(t - \text{ESP} - 2n \cdot \text{ESP})\right],$$

where the ESP is typically a single trapezoid width $T_{\text{trap.}}$, but can also be larger

$$\text{ESP} \geq \tau_{\text{ramp}} + \tau_{\text{flattop}} + \tau_{\text{ramp}} \equiv T_{\text{trap.}}.$$

Two spectral decompositions $g(\omega)$ can be found using the above two infinite train options. Since the main concern here is acoustic energy, the exact phase of $g(\omega)$ can be disregarded and only $|g(\omega)|$ considered, which simplifies the expressions (see SI Appendix S6 for full expressions). Both cases show multiple peaks, or harmonics, with a distance of $2 \cdot \omega_{2\text{ESP}}$ between them — the *odd* harmonics. (Typically, $2 \cdot \omega_{2\text{ESP}} \gg 2\pi/\Delta TE, 2\pi/\Delta T_{\text{slice}}$.) If this distance is large enough, the interaction between the peaks can be neglected and the spectrum around each peak can be approximated separately as

$$|g_n(\omega)| \approx |A_n(\omega_{2\text{ESP}})| \times$$
$$\underbrace{\left|\text{sinc}\left(\tfrac{1}{2}[\omega - \omega_{2\text{ESP}}(2n+1)]T_{\text{ETL}}\right)\right|}_{\text{sinc envelope}} \times \quad \text{(Eq.5)}$$
$$\underbrace{\left|\frac{\sin(\omega \Delta TE N_{\text{TE}}/2)}{\sin(\omega \Delta TE/2)}\right|}_{\text{sinc-like train}} \times \underbrace{\left|\frac{\sin(\omega \Delta T_{\text{slice}} N_{\text{slice}}/2)}{\sin(\omega \Delta T_{\text{slice}}/2)}\right|}_{\text{sinc-like train}} \times \underbrace{\left|\frac{\sin(\omega TR N_{\text{TR}}/2)}{\sin(\omega TR/2)}\right|}_{\text{sinc-like train}},$$

where $|A_n(\omega_{2\text{ESP}})|$ for the sine gradient case is

$$|A_{\text{sine},n}(\omega_{2\text{ESP}})| = \begin{cases} |\pi G \cdot T_{\text{ETL}}| & n = -1, 0 \\ 0 & \text{otherwise} \end{cases},$$

that is only a single harmonic with positive and negative frequencies, while for the trapezoid gradient case $|A_n(\omega_{2\text{ESP}})|$ is

$$|A_{\text{trap.},n}(\omega_{2\text{ESP}})| = \left|2\pi G \frac{\tau_{\text{flattop}} + \tau_{\text{ramp}}}{\text{ESP}} T_{\text{ETL}}\right| \times$$
$$\left|\text{sinc}\left(\left(n + \tfrac{1}{2}\right)\omega_{2\text{ESP}} \cdot \tau_{\text{ramp}}\right) \text{sinc}\left(\left(n + \tfrac{1}{2}\right)\omega_{2\text{ESP}}(\tau_{\text{flattop}} + \tau_{\text{ramp}})\right)\right|, \quad \text{(Eq.6)}$$

that is, an infinite number of (*odd*) harmonics (decaying with $|n|$).

It is interesting to note that in the sinc-like trains above, such as

$$\left|\frac{\sin(\omega \cdot \Delta T_{\text{slice}} \cdot N_{\text{slice}}/2)}{\sin(\omega \cdot \Delta T_{\text{slice}}/2)}\right|,$$

a gradual change of $\Delta T_{\text{slice}}$ to $\Delta T'_{\text{slice}} = \Delta T_{\text{slice}} \pm 2\pi/\omega$ produces, in the vicinity of $\omega$, a shifted-like version of the train until the train practically returns to its original position when $\Delta T'_{\text{slice}} = \Delta T_{\text{slice}} \pm 2\pi/\omega$. In EPI, an $\omega$ around the harmonics, i.e., $\omega \approx \omega_{2\text{ESP}}(2n+1)$, is typically large enough for this behavior to hold.

### Including Navigators in the Acoustic Model

Commonly, EPI scans include an additional short train of three alternating gradient per slice, called the navigators[29]. As no other gradients are on during the navigators, they can be used to estimate inconsistencies between signals measured during positive gradient lobes and during negative gradient lobes. The signals acquired during two opposite lobes should be a time reversal of each other. Any deviation from this between the lobes leads to "ghosts" in the images, as the signal does not match the expected behavior. Measuring the signal during the navigators allows to compensate for such inconsistencies, at least to first order.

The navigators can easily be included in the acoustic model by summing two models, one for the main (longer) gradient train and one for the navigator. For this sum, the complex version of the model (see SI S6 section) should be used.

All other gradients are still disregarded as either their amplitudes are much smaller, or they do not operate at the same ESP cycle (they generally still have a cycle of $\Delta T_{\text{slice}}$).

### Accounting for Mechanical Resonances

The expressions found so far are analytic and do not account for mechanical resonances. These have to be added on top of the analytic model, by multiplying the model by a per-frequency amplification factor, using an appropriate transfer function.

One way of measuring this transfer function is to modify the EPI scan to include only the gradient train, in order for the measured acoustic spectrum to best match the model (while rendering the scan unusable for imaging). Audio recording the scan and dividing the measured acoustic spectrum by the predicted one should give the amplifications due to the mechanical resonances. See Materials and Methods for further details.

## Results

### Effect of Timing on the Analytic Model Spectrum

To examine the effect subtle timing differences can have on the acoustic spectra in multi-echo multi-slice EPI acquisitions, a high temporal resolution case was analyzed. Fig. 1 shows the power spectrum (the spectrum squared) of the trapezoidal-gradient model for an ESP of 0.53 ms, a train length ETL = 54, three TEs, and six slices within a TR of 605 ms. Note that the number of slices in the model actually refers to the number of excitation blocks per TR. Thus, if the simultaneous-multi-slice approach is used, the actual number of slices could be any multiple of the simulated value. The figure shows the power spectrum of a *single* gradient train (single TE and single slice) in yellow; the contribution of the multi-peak factors from the multiple TEs in purple; and the contribution of the slices in green (more closely packed peaks). The product of all three, is the actual power spectrum, in blue. The center-column plots show a case of an "arbitrary" choice of $\Delta TE$ and $\Delta T_{\text{slice}}$, while in the right column $\Delta TE$ and $\Delta T_{\text{slice}}$ are both integer multiples of $2 \cdot \text{ESP}$. The blue power spectra in each case are quite different despite the small change in $\Delta TE$ (0.45 ms) and $\Delta T_{\text{slice}}$ (0.7 ms) between the cases. If $\Delta TE$ and $\Delta T_{\text{slice}}$ are multiples of $2\text{ESP}$, the resulting spectrum consists of a main peak at frequency 1/2ESP and almost negligible peaks in other frequencies, while the 'off 2ESP raster' case creates a collection of substantial peaks with amplitudes dependent on the product of the main squared *sinc* (yellow) with the multiple peaks of the $\Delta TE$ and $\Delta T_{\text{slice}}$ factors (the purple and green peaks).

Note that the changes shown in Fig. 1 do not include the effect of mechanical resonances, which will determine which spectrum is more favorable in practice. SI Appendix Fig. S1 shows another example how different slice timing can significantly affect the spectrum.

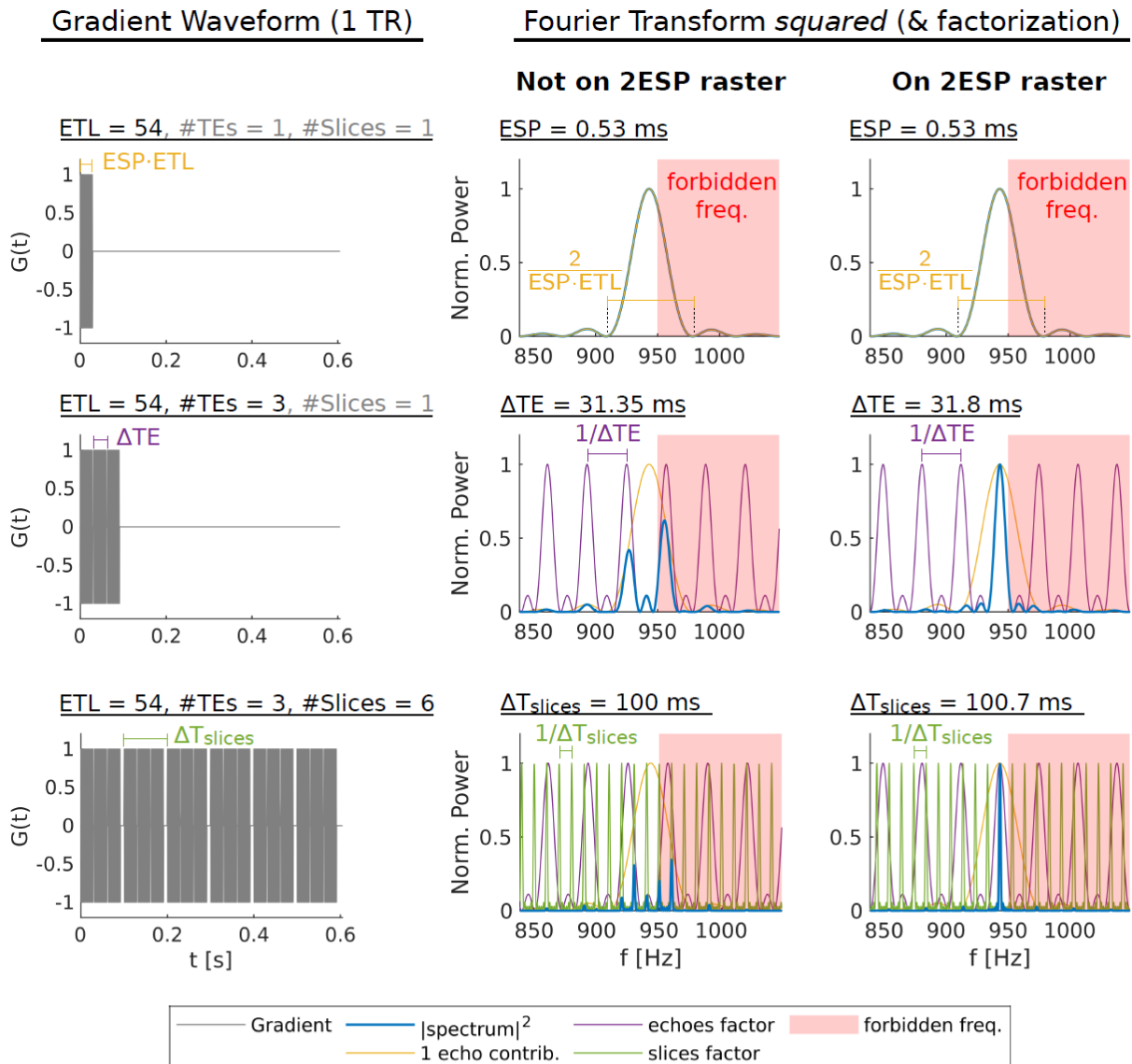

**Fig.1. Modeling acoustic spectrum for a multi-echo multi-slice EPI – demonstrating the effect of subtle timing differences.** Shown are gradient waveforms, made of trapezoidal echo-trains, and their *power* spectrum (the square of the FT), demonstrating the effect of being on and off the 2·ESP raster. Waveforms (top to bottom) include a single slice (single echo-train), 3 echoes and a single slice (3 echo-trains), and three echoes and 6 slices (18 echo-trains). Acoustic spectra (blue) is a numeric FT of the gradient waveforms. For each case, the contributions to the spectrum of a single echo-train (yellow), of multiple echoes (purple), and of multiple slices (green) – whose product forms the full model - are shown. While the plots here show numeric FT for all components, the analytic model delivers the same results (with negligible differences for the first harmonics). Common parameters to all cases are: TR 605 ms, ESP 0.53 ms (0.16 ms ramp up/down and 0.21 ms flat part), and ETL 54.

## Mechanical Resonances and Acoustic Power Prediction

As stated previously, the analytic model is incomplete without the system's actual mechanical resonances which amplify certain frequencies. Here these amplifications, i.e., the transfer function, were measured in a straightforward way, by measuring the actual acoustic spectrum and dividing it by the modeled one. For this, the scan was modified to include only the gradient train (excluding the navigators as well), to best match the model. See Materials and Methods for more details.

The transfer function for three setups was evaluated: the 7T Terra system with the 1Tx/32Rx Nova coil connected, the 7T Terra system with a Flex coil connected (mimics no coil being connected), and the 10.5T MAGNETOM system with a head coil connected. Overall a similar pattern of the transfer functions (SI Appendix Fig. S2) was observed in both the 7T and 10.5T systems (as expected, as both systems have the same type of gradient coils). In all three setups, the profile features higher amplifications in the frequency range restricted by the vendor, as expected. We also observed that the RF coil had a significant effect on the amplification, comparing profiles obtained with the small surface coil and the whole brain head coil. The existing differences between the systems' transfer function will produce modifications in the final acoustics of the scans. Note, that the 10.5T scanner had its third-order shim disconnected, which was recently reported to have a significant impact on the gradient-magnet interactions[10]. This may explain the lower amplification of some peaks, i.e., the disconnection affected the transfer function.

The measured transfer functions were used in conjunction with the analytic model to predict the acoustic power expected from a scan. The predicted acoustic spectrum is the spectrum without the mechanical resonance information multiplied by the estimated "amplification". Summing the squares one can get an estimated (relative) acoustic power.

## Effect of Timing on Measured Acoustic Power and on Ghosting

To experimentally measure the effect of slight timing changes on the acoustics, a product multi-slice 2D EPI scan was modified to enable fine control over the timing of the gradient train and of the navigators. This allowed to sweep through $\Delta T_{\text{slice}}$ and through the navigator's "shift" (increasing its onset delay). In addition, a dual-echo configuration was implemented to evaluate multi-echo and multi-slice 2D EPI (both approaches are of interest for fMRI experiments). For a given EPI set of parameters (including TR, TE, and resolution), acquisition was repeated with varying $\Delta T_{\text{slice}}$ (for multi-slice) and $\Delta TE$ (for dual-echo) in subsets of multi-TR acquisitions (the variation range was usually 0 to 3·ESP). See Materials and Methods for details.

Fig.2A depicts the acoustic spectra resulting from sweeping over the $\Delta T_{\text{slice}}$. Predicted spectra without and with amplification highlight the effect of the timing on the frequency distribution and the amplitude of the peaks. Fig. 2B shows the acoustic power dependence on $\Delta T_{\text{slice}}$ for scan parameters used for mid- and high-temporal resolution acquisition (ESP=0.74 ms and ESP=0.53 ms, respectively). The prediction (scaled to best fit the measured power) matches well the measured behavior, capturing in all cases a distinct cyclic behavior (see Discussion for possible differences). Since these examples exhibit a local peak in the transfer function near the main frequency (1/2ESP), the observed wavelength of the cycle occurs near 2ESP. Varying the slice timing achieves ~2.8-fold difference in the acoustic power at ESP of 0.53 ms for both 7T and 10.5T systems, while at ESP of 0.74 ms ~2-fold difference was achieved. For each scan, a plot of the acoustic spectra for maximal and minimal acoustic energy cases (achieved by varying the $\Delta T_{\text{slice}}$ within ~1 ESP) is also included. One can observe that the spectra peaks are distributed in accordance to the transfer function to appropriately minimize or maximize the energy.

An especially interesting case was observed when the 3$^{rd}$ harmonic of the spectrum was near a mechanical resonance. This is the case with an ESP of 1.26 ms, which is of interest for high spatial resolution fMRI. In this case, Fig.2C , the acoustic peaks around the 3$^{rd}$ harmonic were much higher than the peaks at the 1$^{st}$ harmonic. Cyclic behavior appeared again, however this time the wavelength was ~2·ESP/3 (corresponding to the 3$^{rd}$ harmonics). Both 7T and 10.5T systems show similar results.

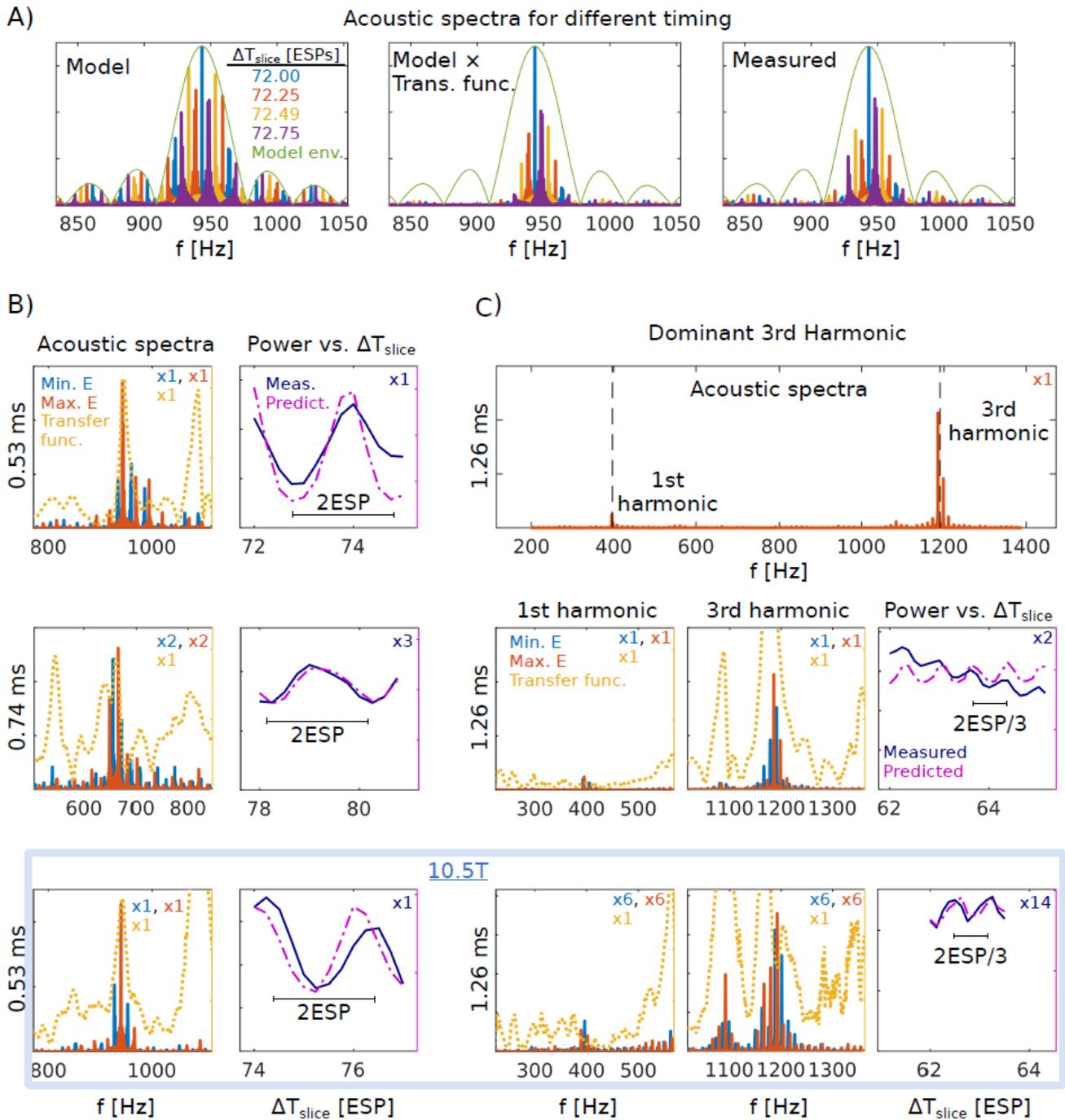

**Fig.2. Acoustic spectra and power, measured and predicted, for subtle slice timing changes.** A) Acoustic spectra for different $\Delta T_{slice}$ – modeled spectra, without and with multiplication by the transfer function, compared to the measured spectra (the green line shows the model of a single echo-train envelope). B) Acoustic power vs $\Delta T_{slice}$ (measured in blue, predicted in purple) for three cases: i) 7T MRI, ESP=0.53 ms, ii) 7T MRI, ESP=0.74 ms, iii) 10.5T MRI, ESP=0.53 ms. C) A case with a dominant 3$^{rd}$ harmonic (ESP=1.26 ms) scanned at 7T and 10.5 T MRI. For each case in B) and C) a maximal power (red) and minimal power (blue) acoustic spectra are shown, as well as the transfer function (dotted yellow). Scaling factors per system (7T or 10.5T) are given in the right corner. The cycle of the power fluctuations is ~2ESP in B), while the case in C) exhibited a cycle of ~2ESP/3, which correlates with the dominant acoustic peak in each case. Note that the predicted power in each case is scaled to best match the measured one (capturing the same relative behavior). See detailed scan parameters in Supporting Information.

Fig.3 highlights the attainable reduction in sound levels in a dual-echo multi-slice EPI configuration when varying both the slice timing and the echo timing. Each dual-echo case is compared to a single-echo multi-slice scan, with variations of both slice timing and navigator timing (for the sake of similarity to the dual-echo case). For example, at an ESP of 0.74 ms, which is commonly used for high resolution imaging, varying both the echo and slice timing achieved a 5-fold difference in measured acoustic power (8-fold in simulations), compared to only a 2-fold difference in the single-echo case. An ESP of 0.4 ms and a short echo-train, of-interest for fast acquisitions, is another interesting case, since its main acoustic peak appears in the vicinity of the highest peak in the transfer function. Here, when scanning a dual-echo multi-slice EPI configuration, a 47-fold difference in the acoustic power was achieved in measurements (57-fold in simulations), while "only" an 8-fold difference was found in the single-echo case. Fig.3 also demonstrates that switching from single-echo to dual-echo EPI – effectively doubling the gradient events per unit time - can yield acoustic power levels comparable to, or even lower than, those of the minimal single-echo condition. Remarkably, at ESP = 0.4 ms, the minimal acoustic power in the dual-echo EPI was reduced by a factor of two compared to the single-echo case.

Our model's predictions show results comparable to the measurements outcomes for all cases (Fig.3). The higher factors predicted in the simulation are likely due to the finer timing resolution in the simulations as well as inaccuracies in the transfer function. Using the model, we also simulated ESP of 0.42 ms, which is 'forbidden' by the system. In this scan, the prediction exhibits 70-fold reduction in acoustic power, to the extent that the sound levels are as low as in non-restricted ESPs (as in ESP of 0.4 ms).

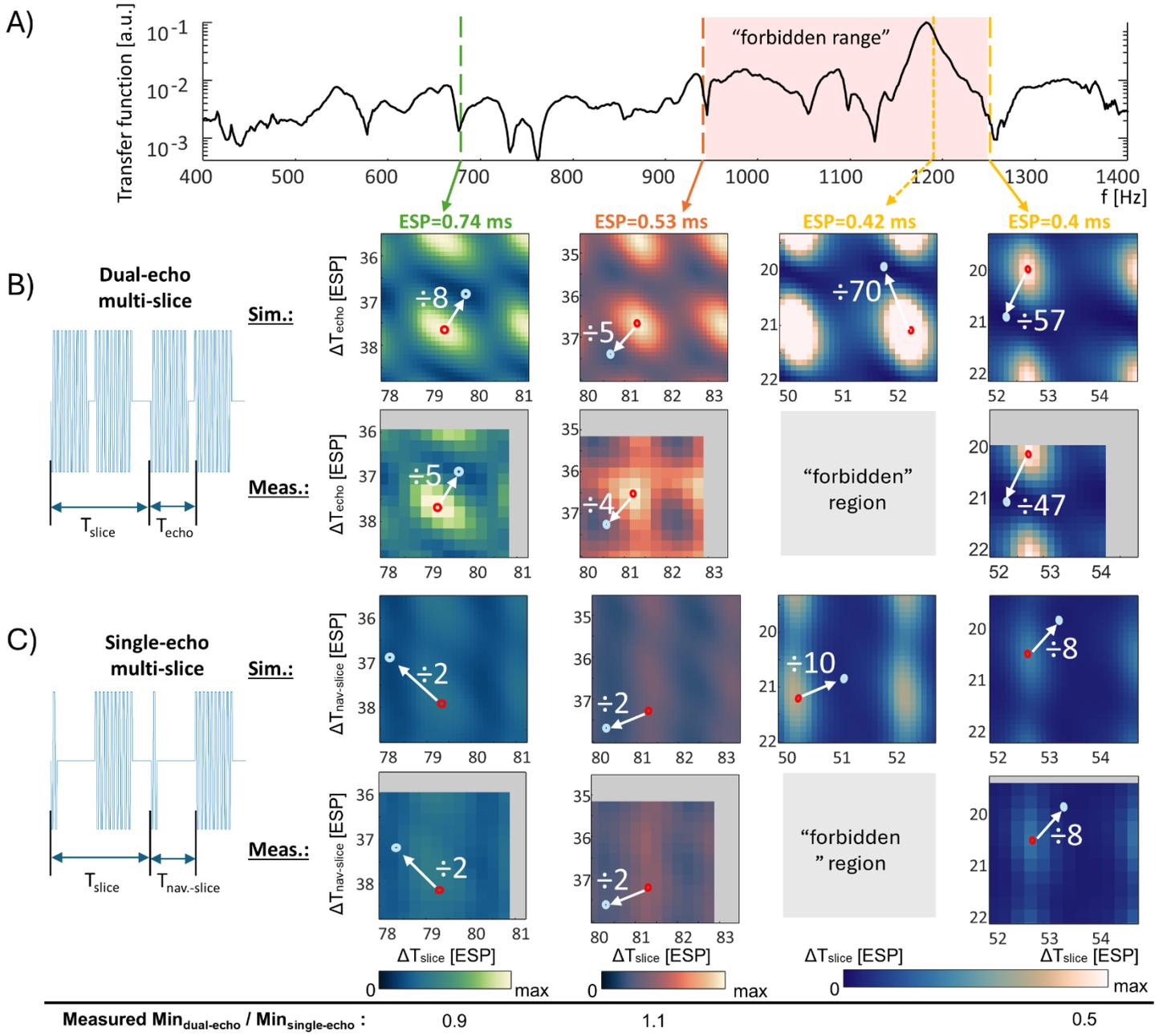

**Fig.3. Measured and predicted acoustic power maps for dual-echo multi-slice EPI when varying echo and slice timing.** A) Transfer function with markings for the 1/2ESP frequency values of the cases shown. B) *Dual*-echo multi-slice EPI - simulated and measured acoustic power maps as function of echo and slice timing. C) *Single*-echo multi-slice EPI - simulated and measured acoustic power maps as function of navigator-to-slice and slice timing. Four sets of scans were examined: ESP = 0.74 ms & ETL=33; ESP=0.53 ms & ETL = 33, ESP=0.42 ms & ETL = 19, and ESP = 0.4 ms & ETL = 19 (See detailed scan parameters in Supporting Information). ESP = 0.42 ms is in the forbidden range, therefore only simulated results are shown. An estimated distance between maxima to maxima is 2ESP for all cases in both directions – echo and slice. For each case the factor from maxima to minima is given in the images. The ratio of the measured minimal acoustic power between dual-echo and single-echo is shown at the bottom.

Fig. 4 illustrates the factors leading to the maximal and minimal power in Fig. 3 for the ESP of 0.4 ms (47-fold measured power change and 57-fold simulated). It shows how the echoes and slice factors, due to the echoes and slice timing, affect the peak locations which are then multiplied by the transfer function.

Since ghosts' artifacts were previously associated with sound levels, we also explored here the ghost level as function of slice and navigator timing. Fig. 5 shows the results of such sweeps for ESP 0.53 ms (at 7T and 10.5T) and for ESP 1.26 ms (at 7T). The scans were performed with a small 4.2 cm diameter ball (see Materials and Methods for more details). Both the effect on the acoustic power and on the ghosting level are shown, as well as sample images with the maximal and minimal ghosting. As can be seen, the navigator timing has little effect on the acoustic power — as might be expected since the navigator has only three alternating gradients. However, the ghosts are clearly affected by the navigator timing. The correlation to the navigator timing has already been shown[30–32], but, here, the connection to vibrations is underlined. As can be seen, the dominant effect on the ghosting is the time from the *end* of one gradient-train to the *start* of the navigator following it. Note that the navigators are used to correct data for the gradient train *following* them, so it is clear that the gradient train *prior* to the navigators is corrupting the navigator data. The results show that a strategically chosen time delay between the navigator and the preceding gradient train can effectively minimize ghosting artifacts. Another observation is that both the acoustic power and the ghosting exhibit similar cyclic patterns. Thus, time delays of the form Δt+n·2ESP (where n is an integer) will also lead to ghost minimization. For the 0.53 ms ESP case, the cycle is approximately 2 · ESP, while for the 1.26 ms ESP, the cycle is approximately 2 · ESP/3. These cycles match the dominant acoustic peaks in each case. The 1$^{st}$ harmonic ($\sim 1/2$ ESP) for the 0.53 ms ESP case, and the 3$^{rd}$ harmonic ($\sim 3/2$ ESP) for the 1.26 ms ESP. The difference between minimal and maximal ghost' level scans reached 5-fold for ESP of 0.53 ms and 3-fold for ESP of 1.26 ms.

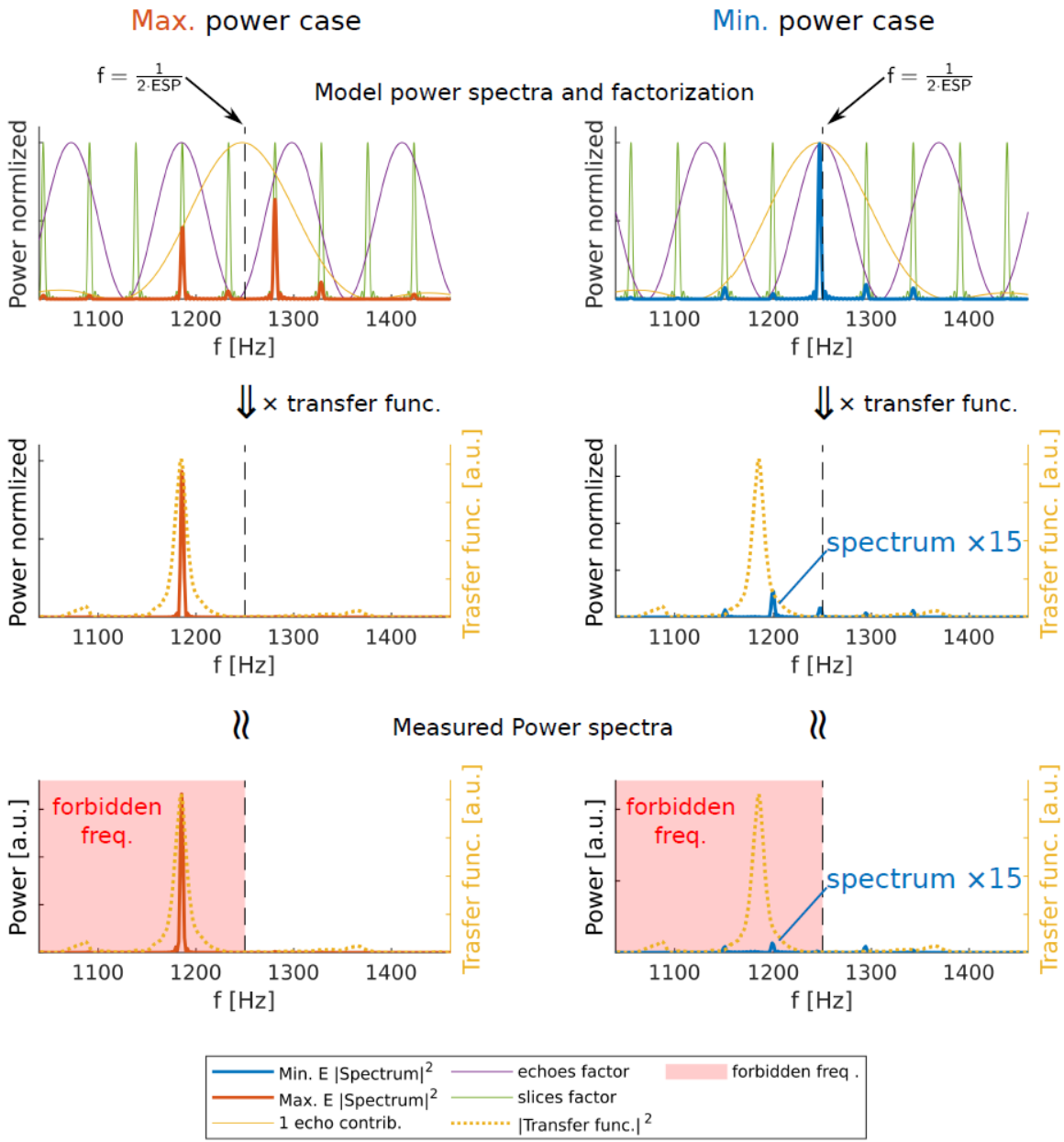

**Fig.4. The source of the difference between maximal and minimal acoustic power cases (Fig.3 ESP=0.4 ms).** Top – Simulated acoustic spectrum for maximal (red, left) and minimal (blue, right) power cases with contributions to the spectrum of a single echo-train (yellow), multiple echoes (purple), and multiple slices (green). Center – the red and blue spectra multiplied by the transfer function (shown as dotted yellow). Bottom – maximal and minimal *measured* power spectra. The difference in timing between the two cases (echo timing difference of 0.4 ms and slice timing difference of 0.23 ms) resulted in different distribution of the purple and green peaks, which as a result generated different final spectra (a product of the yellow, purple and green contributions); multiplying these peaks by the transfer function (dotted-yellow) resulted in one case in a high secondary peak, and in the second case very low secondary peaks. In both cases the main frequency at 1/2ESP was suppressed. It can be seen that in the maximal power case, the measured and simulated spectra match very well, while in the minimal power case, measured and simulated spectra have some deviations. This is because for low peaks, any inaccuracy in the transfer function will affect the result much more.

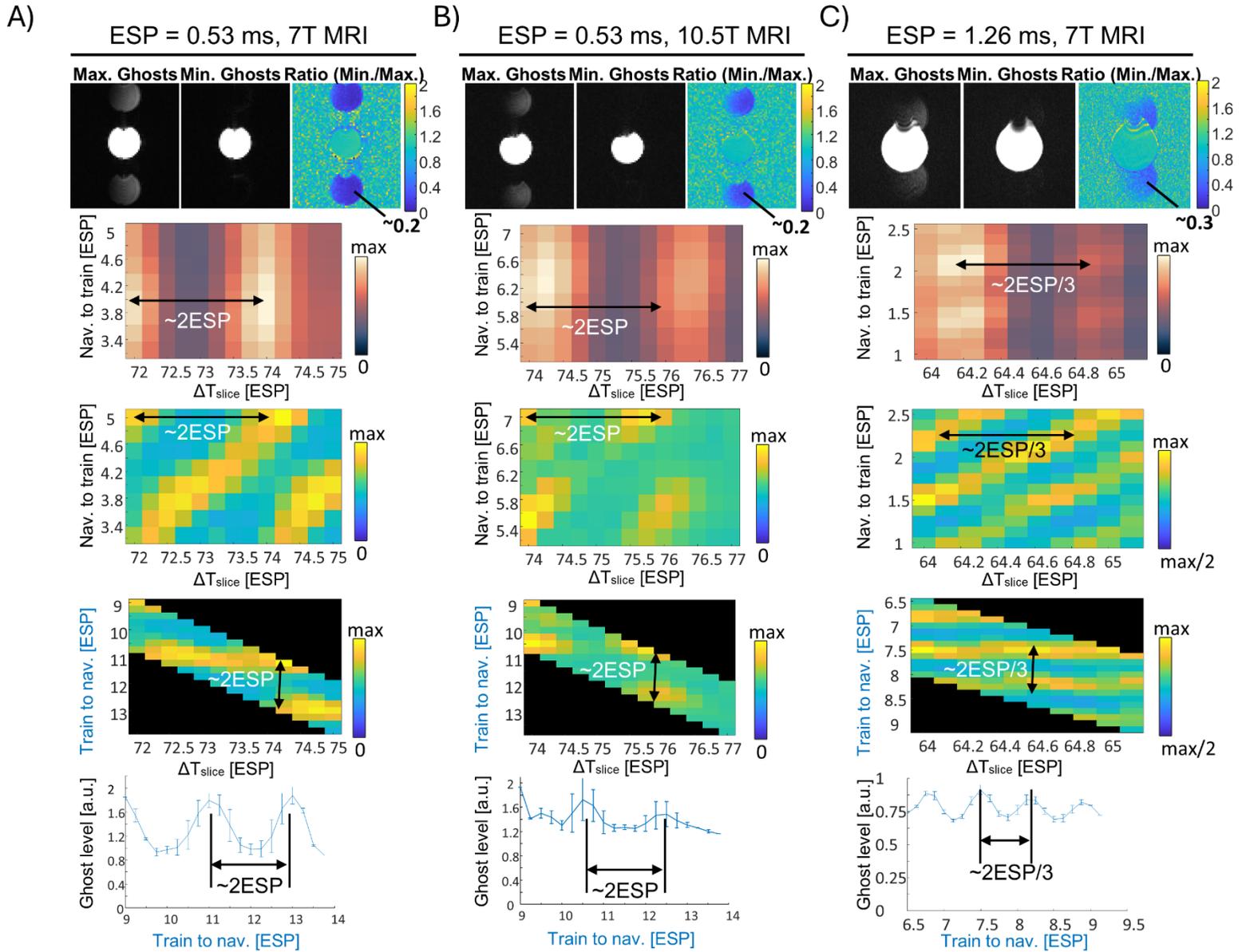

**Fig.5. MRI phantom imaging – effect of subtle timing differences in navigator and slice timing.** A) ESP=0.53 ms, 7T MRI, B) ESP = 0.53 ms, 10.5T MRI, C) ESP = 1.26 ms, 7T MRI. Each case includes, from top to bottom: maximal and minimal ghost images and their ratio; an acoustic power map; an averaged ghost intensity map, as function of $\Delta T_{\text{slice}}$ and navigator-to-echo-train time; the same averaged ghost intensity map, but with y-axis representing the *prior*-echo-train-to-navigator time (instead of the navigator-to-*following*-echo-train time); and the averaged ghost level as function of the prior-echo-train-to-navigator time (error-bars show standard deviation for the different slice times). The cycle of the ghost-level corresponds to the acoustic characteristics (~2ESP for 0.53 ms case and ~2ESP/3 for the 1.26 ms case), either along $\Delta T_{\text{slice}}$ or along the train-to-navigator time direction.

Finally, Fig. 6 shows sample in vivo scans with two ESPs, 0.53 and 1.26 ms. Here the EPI scans repeatedly acquire the volume, as done in "resting state" fMRI. The 0.53 ms ESP case displays ghost-artifacts outside the subject head (due to a field-of-view larger than the brain). The high spatial resolution parameters of the 1.26 ms ESP case limited the field-of-view, so the ghosts occur inside the brain. In both cases, the artifacts were reduced by choosing timing (navigator and slice) based on minimal ghost levels from prior analysis of a phantom (Fig.5). The local signal drops in ESP=1.26 ms are similar to those observed in Ref. [33], where a change of the ESP was used to reduce the artifacts.

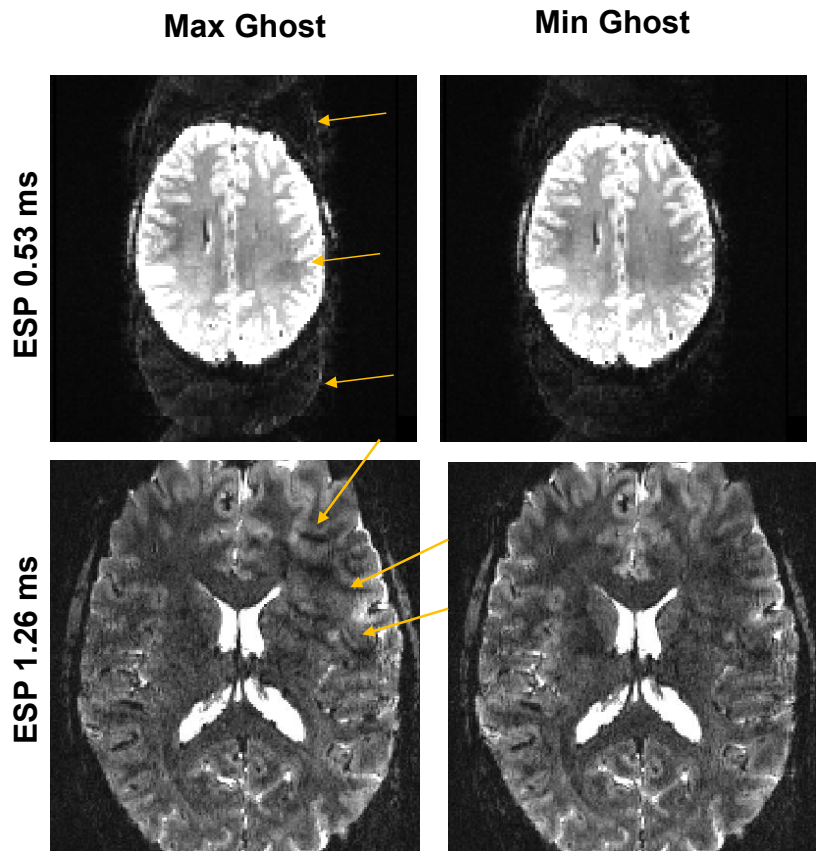

**Fig.6 Human MRI scans using maximal and minimal ghosts-artifact parameters.** Scan parameters were chosen based on phantom data for maximal and minimal ghosts-artifacts for ESP=0.53 ms and ESP=1.26 ms cases. Yellow arrows point to ghosts-artifacts. In the 'maximal ghost' images, the artifacts appear in the first case as ghosts outside the brain and in the second case as local signal drops in the brain.

## Discussion

Two prevalent tenets in EPI are that the chosen ESP determines how loud a scan would be and that loud scans correlate with strong ghosting artifacts. While both tenets are not without basis, it is not the ESP alone that determines the scan sound level nor the ghosting level, but rather an interplay of the ESP, the timing of the gradient trains within the scan, and the system's mechanical resonances. In this study, we introduced a model that can predict and explain how subtle timing changes in EPI can significantly alter mechanical vibration and sound level (reaching measured differences up to ×47 in the acoustic power).

It should be noted that all the results shown here are for gradient trains using the x-gradient coil of the systems. The x-gradient direction was chosen for evaluation as it is the most commonly used in brain fMRI, which typically favors axial or oblique-axial imaging orientations. Similar results are expected using the other gradient coils. The x and y gradient coils are typically very similar (mainly a 90 degree rotation of one another), so similar acoustic transfer functions are expected, however, the z-gradient is different, so a different acoustic transfer function is expected, but otherwise, the same principles should hold.

## Effect of Subtle $\Delta TE$ and $\Delta T_{\text{slice}}$ - Tailoring the Acoustic Spectrum

Considering only the gradient train of the scan while ignoring mechanical resonances, the analytic acoustic spectrum of every harmonic, as given by Eq.5, is basically a sinc envelope of width $\Delta f = 2/T_{\text{ETL}}$ (with $T_{\text{ETL}}$ the duration of the gradient train), multiplied by a product of sinc-like trains (the sine ratios). The interplay of the envelope width and these sinc-like trains with the position and width of the mechanical resonances determines the sound level produced. When the gradient train is sufficiently long — i.e., the width of the sinc envelope is narrower than the mechanical resonance width — the ESP indeed determines which resonance is driven. Similarly, if the envelope sinc is broader, but each sinc-like train (due to $\Delta TE$, $\Delta T_{\text{slice}}$, and $TR$) is closely packed (relative to the mechanical resonance width) then the ESP will still be the dominant parameter as the sinc-like trains will not have much effect — the trains will just discretize the envelope to a resolution finer than the resonance width.

However, if on the one hand the sinc envelope is wide enough while the sinc-like trains have gaps of the order of the mechanical resonance widths, then the exact location of the peaks in the sinc-like trains can have an effect on the final acoustics and sound level, as illustrated in Fig.4. Typically in EPI, the TR is long enough that peaks of its resulting sinc-like train are much more densely packed than the widths of the peaks of either the slice and the TE sinc-like trains, or of the mechanical resonances. Thus, the TR sinc-like train effectively discretizes the spectrum and should have negligible effect. Therefore, the effect of the TR is ignored here.

As noted above, small changes in $\Delta TE$ and $\Delta T_{\text{slice}}$ (as well as in $TR$) appear to cyclically shift each sinc-like train with an apparent cycle, in $\Delta TE$ and $\Delta T_{\text{slice}}$, of $2\pi/\omega$ in the vicinity of any (large enough) $\omega$. When a given angular frequency is dominant with respect to acoustic power (due to the scan timing and mechanical resonance), the power should therefore have a cycle of $2\pi/\omega$ when changing $\Delta TE$ and $\Delta T_{\text{slice}}$. Such a cyclic behavior is indeed observed in Figs. 2,3. The relevant $\omega$ will typically be close to a harmonic of $2\pi/2\,\text{ESP}$, where the dominant harmonic is determined by the system's mechanical resonances. For example, in Fig. 2B (ESPs of 0.53 ms and 0.74 ms), the dominant harmonic is the first one, while in Fig. 2C (ESP of 1.26) the dominant harmonic is the third one. Note that depending on the transfer function, multiple dominant peaks or harmonics can appear (we observed such a case at an ESP of 0.58 ms).

An interesting case occurs when multiple gradient trains are applied per slice, i.e., when $N_{\text{echo}} > 1$. In this case there are two relevant sinc-like trains that interplay, each with its own sinc-like width and sinc-like repetition distance (see Fig.1). Furthermore, $\Delta TE$ in this case is typically very close to the $T_{\text{ETL}}$ duration of a single gradient train, with minimal overhead, so the separation between neighboring peaks in the sinc-like train (purple in Fig. 1) is only slightly less than half the width of the sinc envelope (yellow). This allows a much larger acoustic variation to appear with only very subtle changes to $\Delta TE$ and $\Delta T_{\text{slice}}$, as seen in Figs. 1, 3. In Fig. 3, the measured acoustic power changes for the dual-echo scans are in the range of ×4 to ×47. The ×47 change was achieved by changing $\Delta TE$ by 0.4 ms and $\Delta T_{\text{slice}}$ by 0.23 ms. Importantly, the model can predict this trend, thus demonstrating its potential applicability to the currently forbidden ESP range, where an acoustic power reduction factor of ×70 is predicted, potentially lowering the sound levels to those allowed outside the restricted ESP range. Using this model, the currently "forbidden" echo spacing range of 0.4ms<ESP<0.53ms could be applied in-vivo — for example, in ~2-3 mm resolution protocols — allowing shorter echo-time and, consequently, improved SNR. Such protocols are particularly valuable for imaging regions with short relaxation times, such as the basal ganglia, which play a critical role in studies of movement disorders[34,35].

While the measurements were performed with a dual-echo multi-slice EPI configuration, any combination of multi-echo and multi-slice can now be modeled to develop intuition for possible adaptations. SI Appendix Fig. S5 shows a simulation exploring the acoustic power maps when varying the number of echoes and number of slices for a fixed total number of gradient trains. The figure shows that as the number of echoes (per slice) increases, the area of low acoustic power in the maps expands, making it easier to reduce sound levels with fewer constraints. Notably, the simulation achieved deviations in acoustic power by factors ranging from 20 to 36, contingent on the specific transfer function.

Even with a single gradient train per slice, the effect can be significant. In the single-echo cases examined in Fig. 3, the measured acoustic power dropped by factors of ×2 and ×8, from the highest measured acoustic power to the lowest. This was achieved by applying only subtle timing changes in the slice timing, e.g., changing $\Delta T_{\text{slice}}$ by just ~0.4 ms to achieve an 8-fold reduction.

The timing modifications suggested here are not the only parameters available to tailor the acoustic spectrum. Extending the navigator train and/or the main gradient train will also affect the acoustic spectrum. This is typically feasible in fMRI, as there is often spare time before the EPI's gradient train to achieve the desired contrast. In principle, extending gradients should increase the acoustic power, however, if a mechanical resonance can be avoided this way, then the change may be beneficial. This is demonstrated in Fig. 3 which includes the ratio of the minimal acoustic power between each dual-echo scan and its matching single-echo scan. The ratio is close to 1 in two of the cases, however, in the third case (ESP 0.4 ms), the ratio is 0.5, thus favoring a dual-echo scan despite nearly doubling the gradients per unit time (with otherwise identical parameters).

Changing the sign of some of the gradient trains will also affect the acoustic spectrum, although this should be equivalent to a temporal shift of one ESP of the gradient train, so it could also be applied through a change in $\Delta TE$ or $\Delta T_{\text{slice}}$.

A *Non uniform* distribution of the time *between slices* (and between gradient trains within a slice) is also possible. However, at least in the case of a single gradient train per slice, the tests so far have not shown benefit when using random $\Delta T_{\text{slice}}$, neither in actual experiments nor when predicting the acoustic power (thousands of tries). However, there may be conditions where this is helpful, especially if many slices are acquired per TR.

Previous suggestions to affect the acoustic spectrum focused on changing the gradient train itself. One suggestion was to amplitude modulate the gradient train[12]. This came at a cost of approximately tripling the gradient train, for reconstruction purposes, so that the train's center (where signal is actually acquired) would include almost no amplitude modulation. A second suggestion has been to randomize the ESP within the gradient train[14], here again reconstruction was an issue. It is likely that with current advances, new techniques can better handle the reconstruction; allowing to reduce the overhead needed in the amplitude modulation case and improving the images in the in-train random ESP case. In contrast, the modifications suggested here require no special reconstruction. As long as the changes are maintained in all repetitions (all TRs), each slice is acquired identically every TR, so the nuclear spins within every slice will be in a steady state and no changes are needed to the standard reconstruction.

To summarize, the proposed strategy for minimizing mechanical vibrations of a given scan and reducing acoustic noise can be readily applied in many studies. It is particularly beneficial for fast fMRI applications aiming for minimal echo spacing, including References [20,24,36–39], and multi-echo 2D EPI protocols—such as those used in References [26,27,40–45]. Additionally, any echo spacing that coincides with a local minimum in the acoustic transfer function (e.g., ESP = 0.74 ms as shown in Fig. 3) stands to benefit from this approach by incorporating appropriate timing to reduce acoustic noise. Such local minima are commonly found in the transfer function. Modifying the timing is also relevant for 3D EPI and other EPI variants. It should be noted that in 3D EPI the TR takes the role of $\Delta T_{\text{slice}}$ in 2D EPI, as the gradient train encodes the whole volume each time.

## Predicting Actual Acoustic Power and Acoustic Transfer Function

Including the mechanical resonances in the model requires an acoustic transfer function. By multiplying the model spectrum by the estimated transfer function, the predicted acoustic power successfully captures the oscillations observed in acoustic power (see Figs. 2, 3). However, some deviations still need further analysis. One limitation of the current estimates is that their accuracy is confined to a subset of frequency ranges (~300 Hz – ~1500 Hz and some harmonics). However, this is not a real disadvantage, as the measured acoustic power calculated from this sub-range of frequencies is very similar, mostly a small DC-like reduction in power.

Another factor influencing the transfer function is temperature. It is known that the gradient coil temperature affects the transfer function. In this study, the transfer function was deliberately measured making sure the gradients temperature is low (22°–23° C) to ensure the measurements are consistent. However, during the actual EPI scans, the gradients heat up. SI Appendix Fig. S3 shows the transfer function measured at three different temperature points, illustrating this effect. Employing higher accuracy methods of measuring the transfer function[46–48] can improve the accuracy of the acoustic power prediction.

We also observed that the transfer function changed over time. Appendix Fig. S4 shows repeated measurement of the transfer function over three months. While repeated measurement in a range of a week had negligible deviations, significant deviations were captured over three months. Potential sources of these changes could be slight displacement of the gradient coils due to vibrations, or gradual internal changes within the coils over time, driven by temperature fluctuations and mechanical stress.

Note that with a sufficiently good prediction in hand, one can predict the sound level for patient comfort as well as for better safeguarding of the system from damage and so potentially removing the currently used hard limits on ESPs. As in other protective measures on the scanner — such as specific absorption rate (SAR) and peripheral nerve stimulation — predicting sound levels can be used to supplement a real-time supervision.

## Effect of Subtle $\Delta TE$ and $\Delta T_{\text{slice}}$ – Correlation with Ghosting-Artifacts

As mentioned, a second prevalent tenet is that loud EPI scans correlate with strong ghosting artifacts. While this correlation is not unfounded, the relationship is more complex. It has already been shown[30,32] that the temporal position of the navigators has an effect on the quality of the correction. Here (Fig. 5), we show that the important duration is the time from the end of the gradient train *prior* to the navigators to the start of the navigators. The ghosting-artifact level oscillated according to a cycle matching the frequency of the dominant resonance. This was the same cycle as observed in the acoustic energy case, demonstrating a distinct correlation between acoustics and ghosting-artifacts in the images.

Similar behavior was captured on both 7T and 10.5T MRIs, even though the measurements at 10.5T were performed with the 3rd-order shim disconnected. A previous study has shown that disconnecting the third order shim[10] reduces acoustic sound levels and ghosting artifacts[10,33]. Thus, the effects that were observed in this study are on top of these improvements, showing that ghosting artifacts can be reduced by up to 5-fold through proper timing of the navigator in both systems.

It should also be mentioned that the ghost referred to here are mostly those due to the inconsistencies between positive and negative gradient lobes. However, when acceleration techniques are applied, one also needs to systematically correct for all the reference scans included in the sequence.

## Materials and Methods

### Hardware

Scans were run either on a 7T MRI (MAGNETOM Terra, Siemens Healthcare, Erlangen) or on a 10.5T MRI (MAGNETOM, Siemens Healthineers, Erlangen), with the third order shim of the latter disconnected[10]. Unless otherwise stated all scans at 7T used the 1Tx/32Rx head coil (Nova Medical, Wilmington, MA). On specified scans a 1H TxRx Flex Loop (Rapid Biomedical, Rimpar) was used. On the 10.5T system a custom built 16Tx/80Rx head coil[49] was used.

For audio recordings at the 7T MRI an OptiSLM 100 sound level meter (OptoAcoustics, Mazor) was used. It was connected through a USB/audio interface UCA202 (Behringer, Willich) to a desktop computer where the signal was recorded using Audacity[50]. The microphone was taped at the far end of the bore pointing inwards. On the 10.5T MRI audio was recorded using Bruel and Kjaer Type 2237 SPL meter with Type 4137 Microphone. In this case the microphone was placed at the near end of the bore, pointing inwards.

### Sequence and Scans

For scanning, a modified product EPI sequence was used. The modifications allowed a finer control of the sequence, per slice and per TR, via an external configuration text file. The modifications allowed to control, per TR, the delay after each slice, as well as to delay the start of the navigator. It also allowed to extend the navigator train, adding gradient lobes before and/or after the navigator (signal acquisition was enabled only during the navigator 3-long train). In addition, there was an option to switch off the transmission and/or the gradients (separately per axis) during different blocks within each slice. This allowed to create quiet periods during the scan.

The dual-echo multi-slice EPI was implemented for sound recording only, using the same sequence as above and extending in each slice the navigator echo-train to be of the same length as the acquisition echo train.

The effect of $\Delta T_{\text{slice}}$ and of the navigator timing on the sound level and ghosting was measured using the above scan. The scan was also used to measure the effect of the echo and slice timing on the sound level of the dual-echo case. The scan's configuration text file was setup to step through the navigator delays (or delays between echoes) and then step through the $\Delta T_{\text{slice}}$ values. Each navigator delay and $\Delta T_{\text{slice}}$ configuration was repeated several times in order to reach acoustic steady state, before switching to the next configuration. To help with post analysis, after each sweep of the navigator delay a quiet TR was introduced (switching off the gradient train). Between switching from one $\Delta T_{\text{slice}}$ to the next, two quiet TRs were introduced. Images from the quiet TRs were discarded as the missing gradients produced corrupt images. The scan and sweeping parameters are given in SI Appendix Table S1.

The in-vivo study was approved by the Internal Review Board of the Wolfson Medical Center (Holon, Israel) and all scans were performed after obtaining informed suitable written consents.

## Analysis

All analysis was done using custom Matlab v.9.13 (R2022b) scripts (MathWorks, Natick, MA).

### Audio Recording Analysis

Each audio recording was split into blocks per $\Delta T_{\text{slice}}$ and navigator delay. For this, the quiet blocks built into the scan were used. The split was based on a convolution of the end of the recording (after an initial guess of the end, based on the amplitude recorded) with a boxcar function with the expected length of the non-quiet part (an integer multiple of TR). The convolution maximum was used to mark the location of the block. All other blocks were extracted based on the planned timing of the scan relative to this last non-quiet block.

Once the audio recording was split into blocks, a number of TRs from the end of each block/configuration (typically just the last one) were Fourier transformed to get the measured spectrum — normalized to be independent of repetitions.

Acoustic power was calculated as the sum of squares of the derived (normalized) spectrum. No fixing of units was attempted. As different microphones at different locations were used on the 7T and 10.5T scanners, quantitative comparisons between the systems should not be made.

For the transfer function estimation the measured spectrum and the model spectrum, both in absolute value, were first derived for a single TR, both normalized so that the TR would not effect the scaling. For the measured spectra, the multiple TR repetitions per $\Delta T_{\text{slice}}$ value were used to find average absolute spectra $\left\langle \left| S^{(\text{meas.})}_{\text{ESP},\Delta T_{\text{slice}}}(f) \right| \right\rangle$ and standard deviation of the absolute spectra $\sigma^{(\text{meas.})}_{\text{ESP},\Delta T_{\text{slice}}}(f)$, per frequency.

Here, and elsewhere, the model spectra $\left| S^{(\text{model})}_{\text{ESP},\Delta T_{\text{slice}}}(f) \right|$ were not derived from the analytic expressions, but rather from numeric waveforms of the gradient which were Fourier transformed (after zero padding). The temporal resolution of the waveforms was 10 us to match the scanner's temporal resolution. The resulting spectrum was then interpolated to the measured spectrum frequencies, before applying an absolute value to it. This numeric method was chosen to improve accuracy and speed. To achieve high accuracy with the analytic calculations, many harmonics had to be added together, which proved much slower than performing a numeric calculation.

Once the measured and model spectra were found, the transfer function was found using a weighted average

$$\text{Amplfication}(f) = \sum_{\text{ESP},\Delta T_{\text{slice}}} \frac{\left\langle \left| S^{(\text{meas.})}_{\text{ESP},\Delta T_{\text{slice}}}(f) \right| \right\rangle / \left| S^{(\text{model})}_{\text{ESP},\Delta T_{\text{slice}}}(f) \right|}{\sigma^{(\text{meas.})}_{\text{ESP},\Delta T_{\text{slice}}}(f) / \left| S^{(\text{model})}_{\text{ESP},\Delta T_{\text{slice}}}(f) \right|} \Bigg/ \sum_{\text{ESP},\Delta T_{\text{slice}}} \frac{1}{\sigma^{(\text{meas.})}_{\text{ESP},\Delta T_{\text{slice}}}(f) / \left| S^{(\text{model})}_{\text{ESP},\Delta T_{\text{slice}}}(f) \right|} \quad \text{(Eq.7)}$$

with an estimated standard deviation of

$$\Delta\text{Amplfication}(f) = \sum_{\text{ESP},\Delta T_{\text{slice}}} \frac{1}{\sigma^{(\text{meas.})}_{\text{ESP},\Delta T_{\text{slice}}}(f) / \left| S^{(\text{model})}_{\text{ESP},\Delta T_{\text{slice}}}(f) \right|} . \quad \text{(Eq.8)}$$

Power predictions for specific cases were then found by multiplying the normalized model spectrum for that case by the transfer function, squaring and summing. Because the transfer function was not reliable for the full frequency range, only a subset of frequencies 270Hz – 1500 Hz was used for power estimation. As the transfer function changed over time, each power estimation used the transfer function measured closest to it (same day to a month away).

### Ghost Level Estimation

For ghost level estimation in phantoms, first a mask of the object was found. For this a per pixel (and per slice) "minimum image" was generated from all images in the scan — all $\Delta T_{\text{slice}}$, all navigator delays, and all repetitions — by taking the minimum magnitude value, per pixel, of all cases. This was done assuming the image itself is not much affected, while the ghost will be minimal in this image. After this an initial mask of the object was generated using Matlab's imbinarize() command and holes in the mask were filled using imfill(). The mask was then inverted to include only the area outside the object but then limited to include only pixels within a narrow strip which included the object and ran along the direction the ghosts were expected to appear in (the 'phase encoding' direction). The ghost level measure used was the average signal above the "minimum image" inside this new strip (with a hole in place of the object). Such a measure suffers from a bias due to all the noise pixels without any ghost in them, and misses the ghosting overlapping the object but is simple to implement and is deemed sufficient here, especially in cases where the ghosts do not overlap the object.

## Acknowledgments

We are grateful to Dr. S. Shushan (Wolfson Medical Center) and the Weizmann Institute's MRI technician team - E. Tegareh and N. Oshri - for assistance in the human imaging scans. Dr. R. Schmidt's lab research was generously supported by Mike and Valeria Rosenbloom Center for Research on Positive Neuroscience and Joyce Eisenberg Keefer and Mel Keefer Career Development Chair for New Scientists. Dr. E. Furman-Haran holds the Calin and Elaine Rovinescu Research Fellow Chair for Brain Research. Profs. N. Harel and E. Yacoub were funded by NIH P41EB027061 and U01NS137991.

## Competing interests

A.S and R.S. applied for patents: US Prov. application no.63/777,780 (filed on 26.3.2025) and US Prov. application no.63/841,430 (filed on 10.7.2025).

29. *Handbook of MRI Pulse Sequences*. (Elsevier, 2004). doi:10.1016/B978-0-12-092861-3.X5000-6.

30. Poser B.A., Goa P.E., Barth M. Optimized acquisition strategy for reference-free reduction of Nyquist ghosting in EPI at 7 T. *Proc. Intl. Soc. Mag. Reson. Med. 18 (2010), 5056*.

31. Goa P.E., Jerome N.P. Temporal Oscillation in the Phase Error as an Unresolved Source of Ghosting in EPI at 7T. *Proc. Intl. Soc. Mag. Reson. Med. 29 (2021), 3527*.

32. Poser, B. A., Barth, M., Goa, P., Deng, W. & Stenger, V. A. Single-shot echo-planar imaging with Nyquist ghost compensation: Interleaved dual echo with acceleration (IDEA) echo-planar imaging (EPI). *Magnetic Resonance in Med* **69**, 37–47 (2013).

33. Huber, L. (Renzo) *et al.* Short-term gradient imperfections in high-resolution EPI lead to Fuzzy Ripple artifacts. *Magnetic Resonance in Med* **94**, 571–587 (2025).

34. Puckett, A. M. *et al.* Using multi-echo simultaneous multi-slice (SMS) EPI to improve functional MRI of the subcortical nuclei of the basal ganglia at ultra-high field (7T). *NeuroImage* **172**, 886–895 (2018).

35. de Hollander, G., Keuken, M. C., van der Zwaag, W., Forstmann, B. U. & Trampel, R. Comparing functional MRI protocols for small, iron-rich basal ganglia nuclei such as the subthalamic nucleus at 7 T and 3 T. *Human Brain Mapping* **38**, 3226–3248 (2017).

36. Ushakov, V. L., Orlov, V. A., Kartashov, S. I. & Malakhov, D. G. Ultrafast fMRI sequences for studying the cognitive brain architectures. *Procedia Computer Science* **145**, 581–589 (2018).

37. Sahib, A. K. *et al.* Evaluating the impact of fast-fMRI on dynamic functional connectivity in an event-based paradigm. *PLoS ONE* **13**, e0190480 (2018).
26

# Supporting Information for

Timing is everything: How subtle timing changes in MRI echo planar imaging can significantly alter mechanical vibration and sound level

Amir Seginer[1,2], Alexander Bratch[3], Shahar Goren[2,4], Edna Furman-Haran[1,2], Noam Harel[3], Essa Yacoub[3], Rita Schmidt[2,4]

[1]Life Sciences Core Facilities, Weizmann Institute of Science, Israel

[2]The Azrieli National Institute for Human Brain Imaging and Research, Weizmann Institute of Science

[3]Center for Magnetic Resonance Research (CMRR), University of Minnesota, Minneapolis, Minnesota USA

[4]Department of Brain Sciences, Weizmann Institute of Science, Israel

**This PDF file includes:**

Figures S1 to S6

Tables S1

## S1. Analytic Model

The figure below shows the effect that different distributions of the slices have on the acoustic spectra, within the same TR.

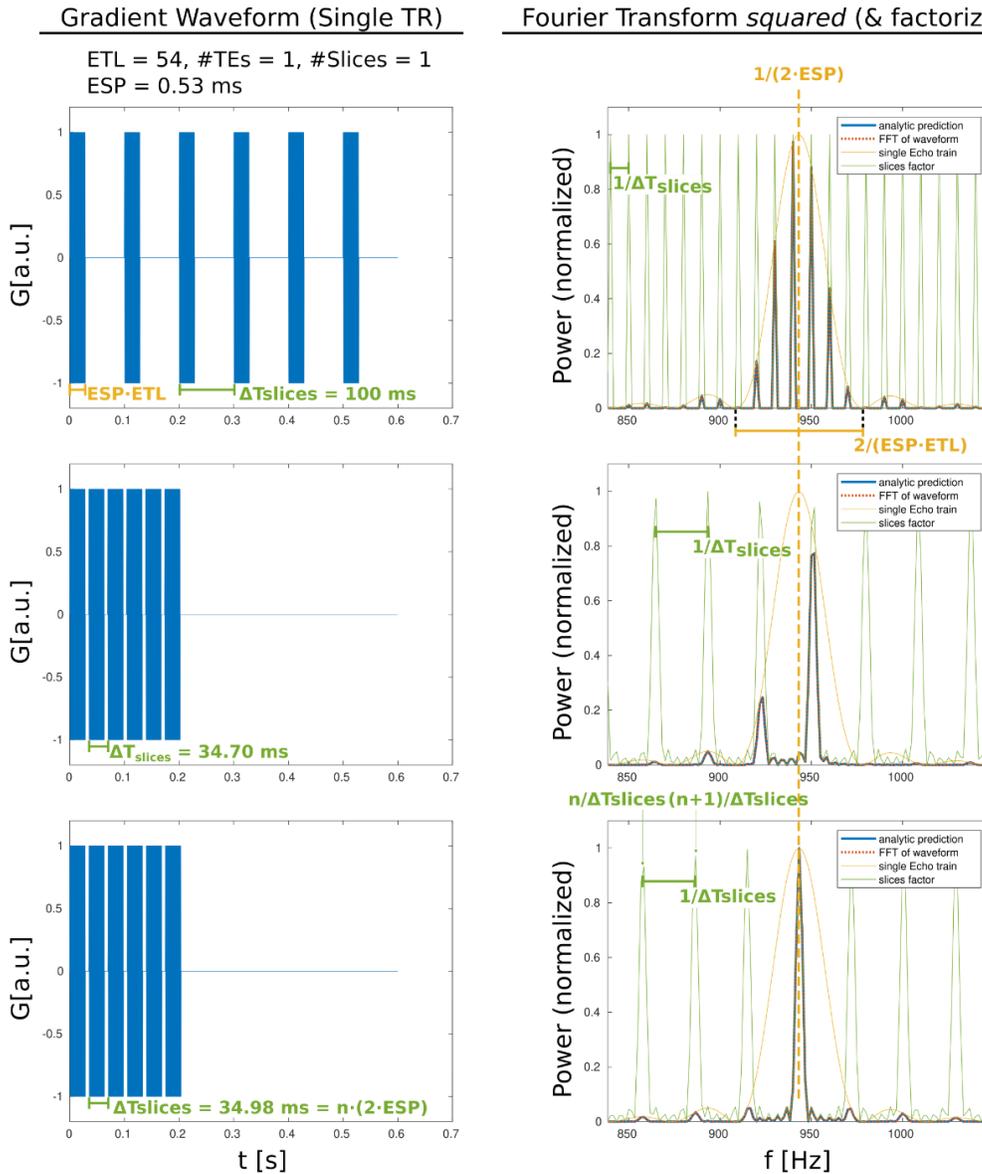

**Fig.S1. Gradient waveforms, made of sinusoidal echo-trains, and their *power* spectrum (the square of the FT), demonstrating the effect of squeezing slices** (6) within a TR and of being on the 2ESP raster. When $\Delta T_{slice}$ is on the 2ESP raster and close to $T_{ETL}$, off-center peaks (away from 1/2ESP) are suppressed. The plots include contributions to the spectrum of a single echo-train (yellow), multiple echos (purple), and multiple slices (green), the product of which is the model (analytic in blue and numeric FFT in dotted red).

## S2. Transfer function

The transfer function of the sound level amplification for each frequency was measured in different setups, as well as a function of temperature and at different time points. The following plots summarize the results. Fig. S2 shows the transfer function measured for three different setups. Overall a similar pattern of the transfer functions was observed in 7T and 10.5T systems (as expected, since both systems have the same type of gradient). In all three setups, the profile features higher amplifications in the frequency range restricted by the vendor, as expected. We also observed that the RF coil had a significant effect on the amplification, comparing profiles obtained with a small surface coil and the whole brain head coil. The differences between the systems' transfer function will produce different acoustics of otherwise identical scans.

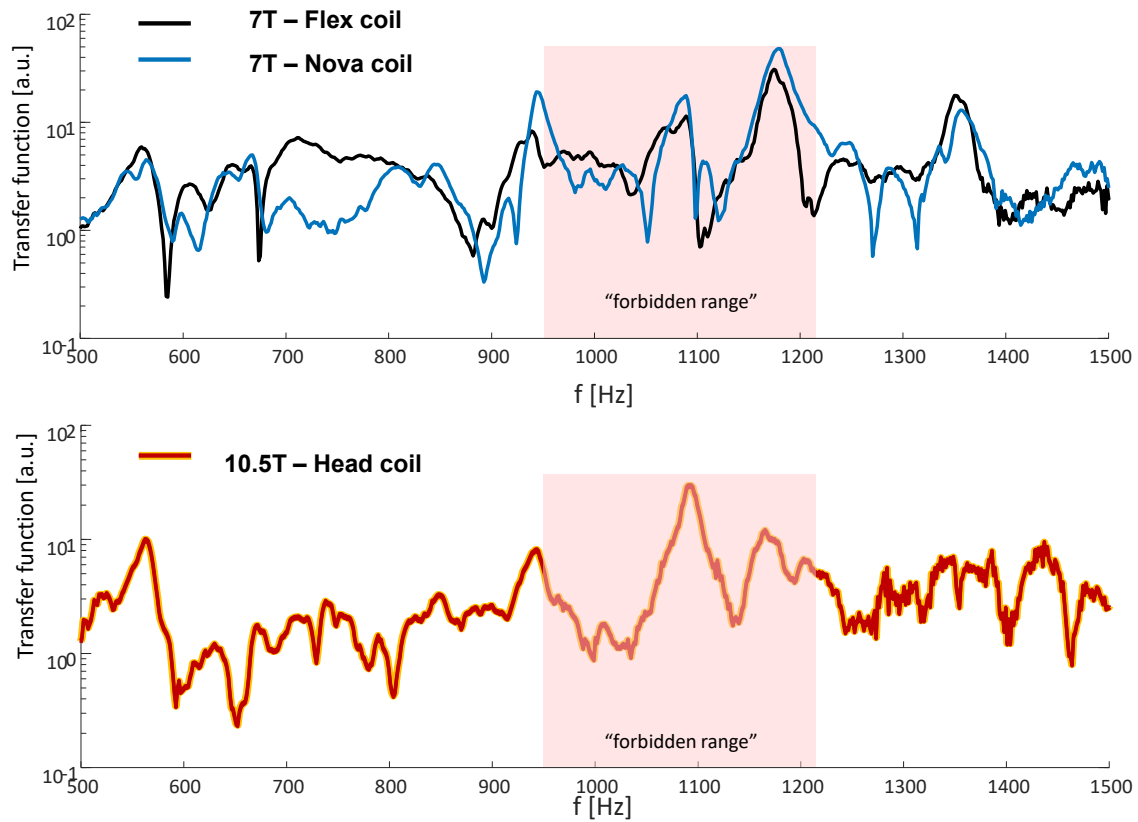

**Fig.S2: Transfer function for 7T and 10.5T MRI setups. For** 7T Terra system the transfer function was evaluated with either the 1Tx/32Rx Nova coil or a Flex coil (mimicing no coil being connected). The 10.5T MAGNETOM system had a head coil connected. Note, that the transfer function of the 10.5T scanner was evaluated with the third-order shim disconnected. For audio recordings at the 7T MRI an OptiSLM 100 sound level meter (OptoAcoustics, Mazor) was used, while on the 10.5T MRI audio was recorded using Bruel and Kjaer Type 2237 SPL meter and Type 4137 Microphone. Therefore, the scaling of the plots was done separately for each system.

To examine transfer function dependence on temperature, we repeated the scans at different temperatures. For that we used the vendor's monitoring of the temperature on the gradients; the monitor with the highest temperature change is reported. Fig. S3 shows the transfer function measured at different temperature points. While for three lower temperature points (21°C, 22°C and 24°C) small deviations in the transfer function were observed, moving to higher temperatures (40°C and 51°C) shows clearly the amplification's dependence on temperature. Note, in some positions the amplification peaks shift with an increase in temperature.

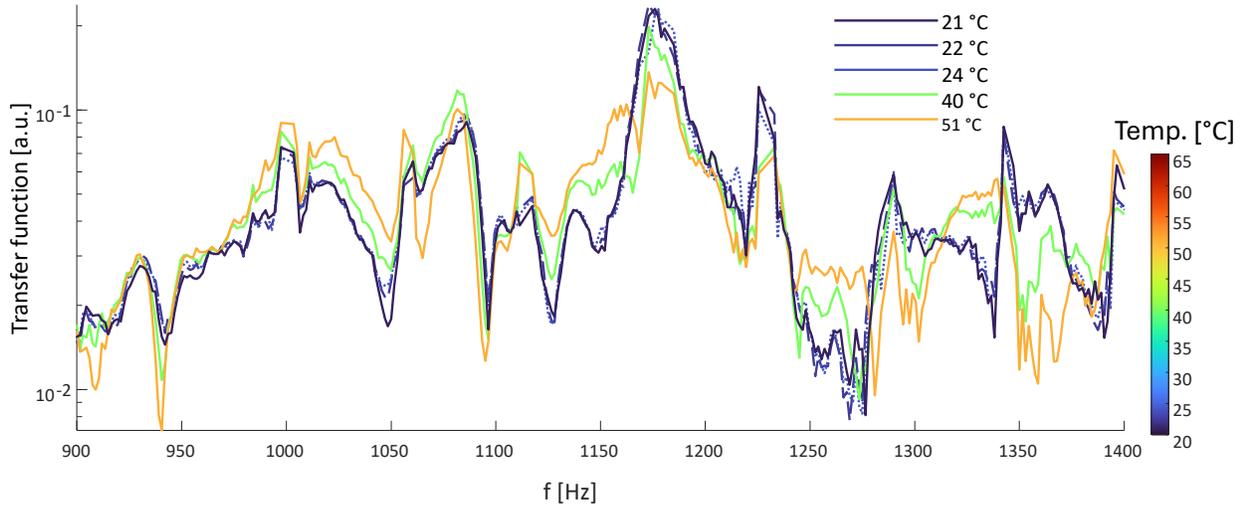

**Fig.S3: Estimated transfer function dependence in temperature.**

We also observed that the transfer function is changed over time. Fig. S4 shows 5 time points over three months.

## S3. Scan parameters

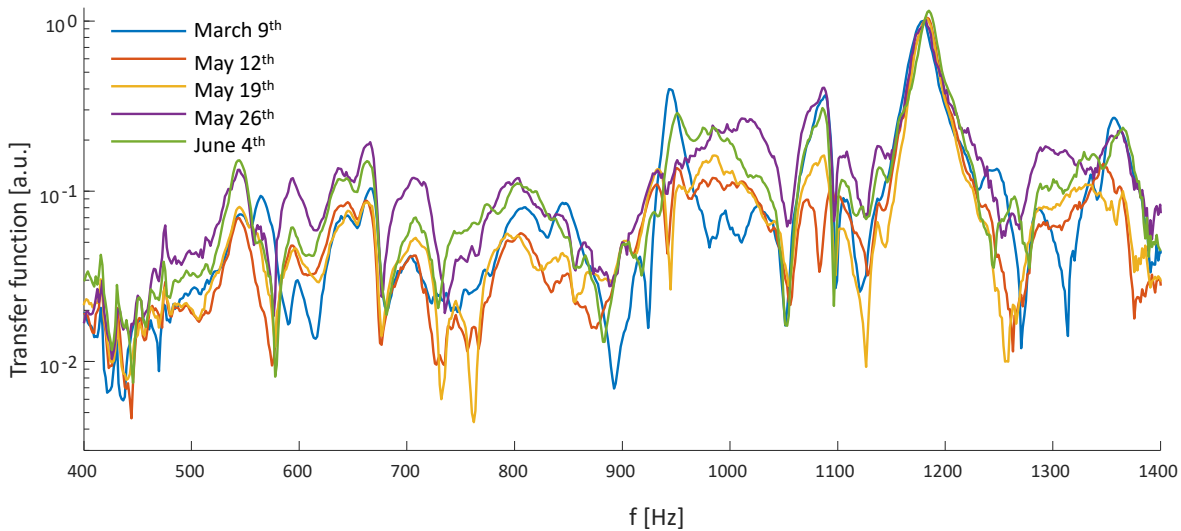

**Fig.S4. Estimated transfer function at different time points over three months.**

The following tables summarize the scan parameters used for the scans whose results are summarized in the figures.

## Table S1, Part 1- Fig.2 scan parameters

|  | Fig.2A | Fig. 2B | Fig. 2B | Fig. 2C |
|---|---|---|---|---|
| Field (B0) [T] | 7 | 7,10.5 | 7 | 7,10.5 |
| ESP [ms] | 0.53 | 0.53 | 0.74 | 1.26 |
| Echo train length (ETL) | 55 | 55 | 33 | 55 |
| TE [ms] | 22 | 21 | 43 | 31 |
| TR [ms] | 900 | 700 | 900 | 3590 |
| Resolution [mm×mm] | 2x2 | 2x2 | 2.2x2.2 | 0.8x0.8 |
| Slice thickness [mm] | 2 | 2 | 2 | 0.8 |
| Field of view (FOV) [mm×mm] | 220x220 | 220x220 | 220x220 | 175x175 |
| Slices | 17 | 15 | 12 | 36 |
| Phase encoding acceleration | 2 | 2 | 3 | 3 |
| Partial Fourier | 1 | 1 | 1 | 6/8 |
| Slice acceleration | 1 | 1 | 1 | 1 |
| $\Delta T_{slice}$ (min./max./#steps) [ms] | 48.76/50.35/13 | 38.16/39.75/13 | 57.72/59.76/12 | 78.12/82.06/26 |
| Navigators delay (min./max./#steps) [ms] | 0.07/1.00/8 | 0.13/1.06/8 | 0.05/2.09/12 | 0.03/0.03/1 |
| TR repetitions per configuration | 5 | 5 | 5 | 3 |

| TRs between $\Delta T_{slice}$ | 2 | 2 | 3 | 2 |
|---|---|---|---|---|
| TRs between navigator delays | 1 | 1 | 2 | 1 |

## Table S1, Part 2 - Fig.3 scan parameters

| Field (B0) [T] | 7 | 7 | 7 |
|---|---|---|---|
| ESP [ms] | 0.74 | 0.53 | 0.4 |
| Echo train length (ETL) | 33 | 33 | 19 |
| TE [ms] | 43 | 31 | 15 |
| TR [ms] | 900 | 700 | 680 |
| Resolution [mm×mm] | 2.2x2.2 | 2.2x2.2 | 6x6 |
| Slice thickness [mm] | 2 | 2 | 2 |
| Field of view (FOV) [mm×mm] | 220x220 | 220x220 | 383x500 |
| Slices | 12 | 12 | 15 |
| Phase encoding acceleration | 3 | 3 | 2 |
| Partial Fourier | 1 | 1 | 1 |
| Slice acceleration | 1 | 1 | 1 |
| $\Delta T_{slice}$ (min./max./#steps) [ms] | 57.72/59.76/12 | 42.40/43.86/12 | 20.80/21.90/12 |
| Navigators delay (min./max./#steps) [ms] | 0.05/2.09/12 | 0.06/1.52/12 | 0.00/1.10/12 |

| TR repetitions per configuration | 5 | 5 | 5 |
| --- | --- | --- | --- |
| TRs between $\Delta T_{\text{slice}}$ | 3 | 3 | 3 |
| TRs between navigator delays | 2 | 2 | 2 |

Figs.5 and 6 parameters for ESP 0.53 ms and ESP 1.26 ms were the same as in Fig.2B for ESP 0.53 ms and Fig.2C for ESP 1.26 ms.

## S4. Multi-echo multi-slice simulation

A simulation was conducted in which the number of echoes and number of slices was varied while keeping the total number of gradient trains fixed. The scan parameters were: TR = 701 ms, TE = 22 ms, ETL = 55, ESP = 0.53 ms, and number of slices = 15. Fig.S5 shows the resulting acoustic power as function of $\Delta T_{slice}$ and $\Delta TE$.

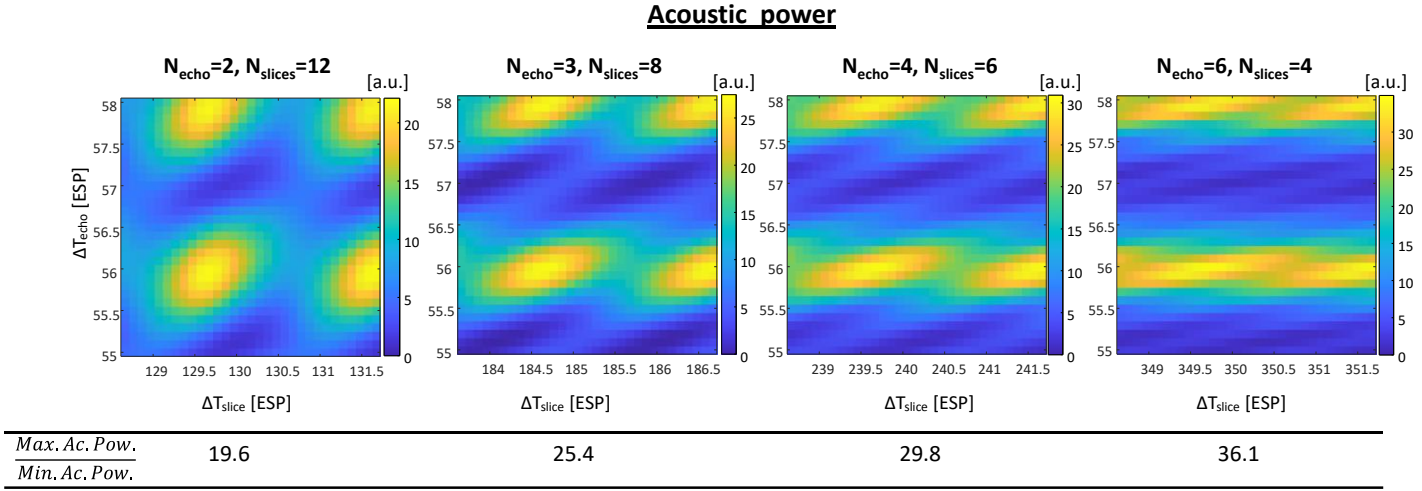

| $\frac{Max.Ac.Pow.}{Min.Ac.Pow.}$ | 19.6 | 25.4 | 29.8 | 36.1 |

**Fig.S5.** A simulation exploring the acoustic power maps when varying the number of echoes and slices for a fixed total of gradient trains.

## S5. Pulse sequence diagram

Fig.S6 shows a multi-slice 2D EPI pulse sequence diagram (two slices) illustrating the $\Delta T_{slice}$ and navigator shift definitions.

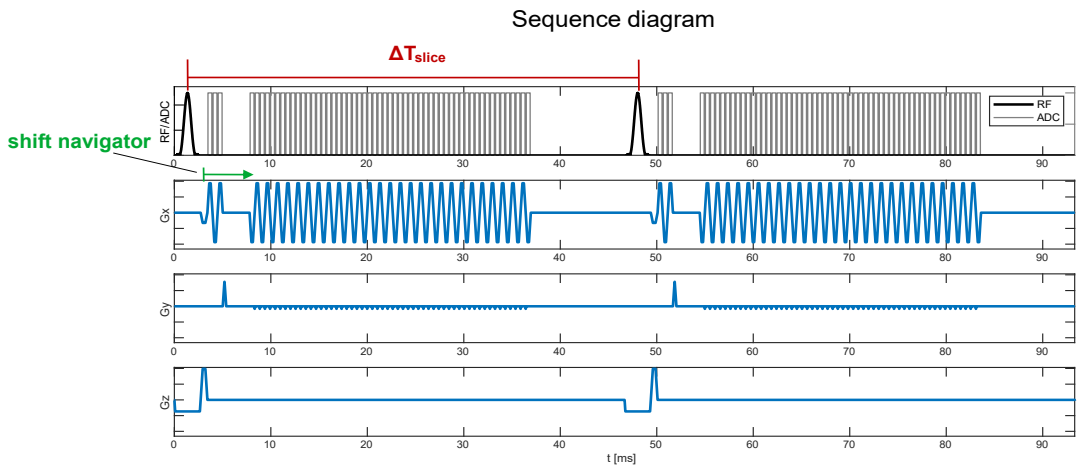

**Fig.S6.** Multi-slice 2D EPI pulse sequence diagram with definition of the the $\Delta T_{slice}$ and

# S6. Model Derivation

## S6.1. Introduction

As mentioned in the main text the basic acoustic spectrum model used here (disregarding mechanical resonances) is the Fourier transform of the alternating gradients used in EPI. Thus to derive the model, first a gradient waveform $G(t)$ has to be defined and then its Fourier transform has to be determined.

More complicated models can be generated by either using a more complicated gradient waveform $G(t)$, or by combining simple waveforms to create the complicated $G(t)$ which is equivalent to adding together spectra from simple waveforms.

The model constructed here is for multi-slice EPI, where each slice is acquired separately. However, it should easily accommodate 3D EPI, where all the volume is acquired together (with an extra Fourier dimension), as long as the train length is kept constant within the acquisition.

## S6.2. The Gradient Waveform

The gradient waveform $G(t)$ can be constructed in steps. The first is to generate a single gradient train $G^{(\text{ETL})}(t)$ of length ETL (echo train length). Then a single slice may include multiple ($N_{\text{TE}}$) such trains, termed echoes,[1] every $\Delta TE$ time interval (i.e., $\Delta TE$ from the start of one to the start of the next). We shall add the option for the sign of the gradient trains $G^{(\text{ETL})}(t)$ to alternate between echoes. Once the gradient of a single slice is defined, the set of all slices is just a repeated set ($N_{\text{slice}}$-times) of such gradients, every $\Delta T_{\text{slice}}$ time interval. Finally, the set of slices can be repeated $N_{\text{TR}}$ times, in this case every $TR$ time interval.

Two types of gradients are considered here, a sine-shaped gradient train and a trapezoid shaped gradient train.

### S6.2.1. Sine-Shaped Single Gradient Train

A single sine-shaped gradient train with a cycle of two echo-spacings (ESP) can be viewed as a short section out of an infinite sine-shaped train, or

$$G_{\text{sine}}^{(\text{ETL})}(t) = G_{\text{sine}}^{(\infty)}(t) \cdot \Pi(t/T_{\text{ETL}}), \tag{S1}$$

where $G_{\text{sine}}^{(\infty)}(t)$ is an infinite sine-shaped gradient

$$G_{\text{sine}}^{(\infty)} \equiv G \sin(\omega_{2\text{ESP}} t) \tag{S2}$$

with

$$\omega_{2\text{ESP}} \equiv 2\pi f_{2\text{ESP}} \equiv \frac{2\pi}{2\text{ESP}}. \tag{S3}$$

And where $\Pi(t)$ is the boxcar function, defined here as

$$\Pi(t) \equiv \begin{cases} 1 & 0 < t < 1 \\ 0 & \text{otherwise} \end{cases}, \tag{S4}$$

and $T_{\text{ETL}}$ is the duration of the gradient train

$$T_{\text{ETL}} \equiv \text{ETL} \cdot \text{ESP}. \tag{S5}$$

---

[1] As in the main text, note that the multiple echoes within the slice are not the same echoes as in the 'echo train length'. The latter are within the gradient train, while the prior are different trains.

### S6.2.2. Trapezoid-Shaped Single Gradient Train

When the basic unit is a trapezoid, we limit ourselves to *symmetric* trapezoids. A convenient definition, for the purposes here, of a single symmetric trapezoid of amplitude $G$, ramp up/down times of $\tau_{\text{ramp}}$, and a flat part of duration $\tau_{\text{flattop}}$ is to define it as a convolution of two boxcars as follows

$$G^{(1)}_{\text{trap.}} = \frac{G}{\tau_{\text{ramp}}} \Pi(\frac{t}{\tau_{\text{ramp}}}) * \Pi(\frac{t}{\tau_{\text{flattop}} + \tau_{\text{ramp}}}) \quad \text{(single trapezoid).} \tag{S6}$$

To be consistent with the sine-shaped train above, an infinite gradient train with a cycle of $2\text{ESP}$ can now be defined as

$$G^{(\infty)}_{\text{trap.}}(t) = \overbrace{\left[\frac{G}{\tau_{\text{ramp}}} \Pi(\frac{t}{\tau_{\text{ramp}}}) * \Pi(\frac{t}{\tau_{\text{flattop}} + \tau_{\text{ramp}}})\right]}^{G^{(1)}_{\text{trap.}}} * \\ \left[\sum_{n=-\infty}^{\infty} \delta(t - 2n\text{ESP}) - \sum_{n=-\infty}^{\infty} \delta(t - \text{ESP} - 2n\text{ESP})\right], \tag{S7}$$

which is just a convolution of $G^{(1)}_{\text{trap.}}(t)$ with an infinite train of alternating delta functions (one ESP between positive and negative delta functions).

A single gradient train is then just, as before, a section of this infinite train

$$G^{(\text{ETL})}_{\text{trap.}} = G^{(\infty)}_{\text{trap.}}(t) \cdot \Pi(\frac{t}{T_{\text{ETL}}}). \tag{S8}$$

### S6.2.3. Single Slice Gradient Waveform

Once we have an expression for a single gradient train $G^{(\text{ETL})}(t)$, a single slice simply includes $N_{\text{TE}}$ echoes every $\Delta TE$, starting from a time $T_{0,\text{TE}}$, which is simply the convolution of $G^{(\text{ETL})}(t)$ with another set of delta functions

$$G^{(\text{slice})}(t) = G^{(\text{ETL})}(t) * \sum_{n_{\text{TE}}=0}^{N_{\text{TE}}-1} e^{i\theta \cdot n_{\text{TE}}} \delta(t - T_{0,\text{TE}} - n_{\text{TE}}\Delta TE), \tag{S9}$$

where $\theta$ is either zero or $\pi$, so the echoes are all the same ($\theta = 0$), or alternate in sign ($\theta = \pi$).

### S6.2.4. Single TR Gradient Waveform

A single volume or repetition is made up of $N_{\text{slice}}$ slices every $\Delta T_{\text{slice}}$, starting from a time $T_{0,\text{slice}}$, which is again a convolution of the previous step, $G^{(\text{slice})}(t)$, with a new set of delta functions

$$G^{(\text{TR})}(t) = G^{(\text{slice})}(t) * \sum_{n_{\text{slice}}=0}^{N_{\text{slice}}-1} \delta(t - T_{0,\text{slice}} - n_{\text{slice}}\Delta T_{\text{slice}}). \tag{S10}$$

### S6.2.5. Repeated Acquisitions Gradient Waveform

Finally, repeating the acquisition $N_{\text{TR}}$ times every $TR$, the last convolution is of $G^{(\text{TR})}(t)$ with a set of delta functions every $TR$. This gives the final gradient waveform

$$G(t) = G^{(\text{TR})}(t) * \sum_{n_{\text{TR}}=0}^{N_{\text{TR}}-1} \delta(t - n_{\text{TR}}TR). \tag{S11}$$

### S6.2.6. Final Gradient Waveform

Combining equations (S9)–(S11), one can write

$$G(t) = G^{(\text{ETL})}(t) * \quad (S12)$$

$$\sum_{n_{\text{TE}}=0}^{N_{\text{TE}}-1} e^{i\theta \cdot n_{\text{TE}}} \delta(t - T_{0,\text{TE}} - n_{\text{TE}}\Delta TE) *$$

$$\sum_{n_{\text{slice}}=0}^{N_{\text{slice}}-1} \delta(t - T_{0,\text{slice}} - n_{\text{slice}}\Delta T_{\text{slice}}) *$$

$$\sum_{n_{\text{TR}}=0}^{N_{\text{TR}}-1} \delta(t - n_{\text{TR}} TR),$$

where $G^{(\text{ETL})}(t)$ is either Eq. (S1) for sine-shaped gradient case, or Eq. (S8) for the trapezoid-shaped gradient. Using these expressions two forms for the gradient waveforms can be written.

For the sine-shaped case

$$G_{\text{sine}}(t) = \left[ \overbrace{G \sin(\omega_{2\text{ESP}} t) \cdot \Pi(t/T_{\text{ETL}})}^{G_{\text{sine}}^{(\infty)}(t)} \right] * \quad (S13)$$

$$\sum_{n_{\text{TE}}=0}^{N_{\text{TE}}-1} e^{i\theta \cdot n_{\text{TE}}} \delta(t - T_{0,\text{TE}} - n_{\text{TE}}\Delta TE) *$$

$$\sum_{n_{\text{slice}}=0}^{N_{\text{slice}}-1} \delta(t - T_{0,\text{slice}} - n_{\text{slice}}\Delta T_{\text{slice}}) *$$

$$\sum_{n_{\text{TR}}=0}^{N_{\text{TR}}-1} \delta(t - n_{\text{TR}} TR).$$

While for the trapezoid-shaped gradient, the form consistent with the sine-shaped case (using an infinite train) is

$$G_{\text{trap.}}(t) = \left\{ \left[ \frac{G}{\tau_{\text{ramp}}} \Pi(t/\tau_{\text{ramp}}) * \Pi(t/(\tau_{\text{flattop}} + \tau_{\text{ramp}})) \right] * \quad (S14) \right.$$

$$\left. \left[ \sum_{n=-\infty}^{\infty} \delta(t - 2n\text{ESP}) - \sum_{n=-\infty}^{\infty} \delta(t - \text{ESP} - 2n\text{ESP}) \right] \cdot \Pi(t/T_{\text{ETL}}) \right\} *$$

$$\sum_{n_{\text{TE}}=0}^{N_{\text{TE}}-1} e^{i\theta \cdot n_{\text{TE}}} \delta(t - T_{0,\text{TE}} - n_{\text{TE}}\Delta TE) *$$

$$\sum_{n_{\text{slice}}=0}^{N_{\text{slice}}-1} \delta(t - T_{0,\text{slice}} - n_{\text{slice}}\Delta T_{\text{slice}}) *$$

$$\sum_{n_{\text{TR}}=0}^{N_{\text{TR}}-1} \delta(t - n_{\text{TR}} TR).$$

## S6.3. The Spectrum (Post Fourier Transform)

### S6.3.1. The Fourier Transform

As stated earlier, the spectrum $g(\omega)$ is the Fourier transform of the gradient waveform $G(t)$. For this, the Fourier transform used here is defined as

$$g(\omega) = \mathcal{F}\{G(t)\} \equiv \int_{-\infty}^{+\infty} e^{-i\omega t} G(t) dt, \tag{S15}$$

with an inverse

$$\mathcal{F}^{-1}\{g(\omega)\} = \frac{1}{2\pi} \int_{-\infty}^{+\infty} e^{i\omega t} g(\omega) d\omega. \tag{S16}$$

We shall also make use of the convolution theorem. The convolution is defined as

$$F(t) * G(t) \equiv \int_{-\infty}^{+\infty} F(t')G(t-t')dt', \tag{S17}$$

and the convolution theorem states that

$$\mathcal{F}\{F(t) * G(t)\} = \mathcal{F}\{F(t)\} \cdot \mathcal{F}\{G(t)\}, \tag{S18}$$

as well as

$$\mathcal{F}\{F(t) \cdot G(t)\} = \frac{1}{2\pi} \mathcal{F}\{F(t)\} * \mathcal{F}\{G(t)\}. \tag{S19}$$

The Fourier transform of a single delta function is

$$\mathcal{F}\{\delta(t-t_0)\} = e^{-i\omega t_0},$$

but the Fourier transform of an infinite series of delta functions is also an infinite set of delta functions

$$\mathcal{F}\left\{\sum_{n=-\infty}^{\infty} \delta(t - \tau_0 - n\tau)\right\} = \frac{2\pi}{\tau} e^{-i\omega \tau_0} \sum_{n=-\infty}^{\infty} \delta\left(\omega - \frac{2\pi}{\tau} n\right), \tag{S20}$$

and for a finite series of delta functions (using the sum of a finite geometric series)

$$\mathcal{F}\left\{\sum_{n=0}^{N-1} e^{i\theta n} \delta(t - \tau_0 - n\tau)\right\} = e^{-i\omega \tau_0} e^{-i(\omega-\omega_\theta)\tau(N-1)/2} \frac{\sin\left((\omega-\omega_\theta)\tau N/2\right)}{\sin\left((\omega-\omega_\theta)\tau/2\right)}, \tag{S21}$$

where

$$\omega_\theta \equiv \theta/\tau. \tag{S22}$$

We will also need the Fourier transform of a boxcar function, Eq. (S4), which is just

$$\mathcal{F}\{\Pi(t/\tau)\} = \tau e^{-i\omega \tau/2} \operatorname{sinc}\left(\frac{1}{2}\omega\tau\right), \tag{S23}$$

and the Fourier transform

$$\mathcal{F}\left\{e^{\pm i\omega_0 t}\right\} = 2\pi \delta(\omega \mp \omega_0). \tag{S24}$$

### S6.3.2. Deriving the Spectrum

Starting from the general Eq. (S12) and using the convolution theorem (S18) as well as the Fourier transform of a finite series (S21), we can write

$$g(\omega) = \mathcal{F}\{G(t)\} \tag{S25}$$
$$= \mathcal{F}\left\{G^{(\text{ETL})}(t)\right\} \times$$
$$e^{-i\omega T_{0,\text{TE}}} e^{-i(\omega-\omega_\theta)\Delta TE(N_{\text{TE}}-1)/2} \frac{\sin\left((\omega-\omega_\theta)\Delta TE \cdot N_{\text{TE}}/2\right)}{\sin\left((\omega-\omega_\theta)\Delta TE/2\right)} \times$$
$$e^{-i\omega T_{0,\text{slice}}} e^{-i\omega\Delta T_{\text{slice}}(N_{\text{slice}}-1)/2} \frac{\sin\left(\omega\Delta T_{\text{slice}} \cdot N_{\text{slice}}/2\right)}{\sin\left(\omega\Delta T_{\text{slice}}/2\right)} \times$$
$$e^{-i\omega TR(N_{\text{TR}}-1)/2} \frac{\sin\left(\omega TR \cdot N_{\text{TR}}/2\right)}{\sin\left(\omega TR/2\right)}.$$

It now remains to find $\mathcal{F}\left\{G^{(\text{ETL})}(t)\right\}$ for the different cases. In the case of a sine-shaped gradient, using Eqs. (S1), (S2), (S19), (S23), (S24) and using

$$\sin \omega t = \frac{e^{i\omega t} - e^{-i\omega t}}{2i} \tag{S26}$$

it is

$$\mathcal{F}\left\{G^{(\text{ETL})}_{\text{sine}}(t)\right\} = \mathcal{F}\{G\sin(\omega_{2\text{ESP}}t)\} * \mathcal{F}\{\Pi(t/T_{\text{ETL}})\} \tag{S27}$$
$$= -i\pi GT_{\text{ETL}} \left[ e^{-i(\omega-\omega_{2\text{ESP}})T_{\text{ETL}}/2} \operatorname{sinc}\left(\frac{1}{2}(\omega-\omega_{2\text{ESP}})T_{\text{ETL}}\right) - \right.$$
$$\left. e^{-i(\omega+\omega_{2\text{ESP}})T_{\text{ETL}}/2} \operatorname{sinc}\left(\frac{1}{2}(\omega+\omega_{2\text{ESP}})T_{\text{ETL}}\right) \right].$$

Now, for the trapezoid case, using Eqs. (S7) and (S8) with (S18), (S19), and (S23), the Fourier transform $\mathcal{F}\left\{G^{(\text{ETL})}(t)\right\}$ in this case is

$$\mathcal{F}\left\{G^{(\text{ETL})}_{\text{trap.}}(t)\right\} = \left[\mathcal{F}\left\{\frac{G}{\tau_{\text{ramp}}}\Pi(t/\tau_{\text{ramp}})\right\} \cdot \mathcal{F}\left\{\Pi\left(t/(\tau_{\text{flattop}}+\tau_{\text{ramp}})\right)\right\} \times \right. \tag{S28}$$
$$\left. \mathcal{F}\left\{\sum_{n=-\infty}^{\infty}\delta(t-2n\text{ESP}) - \sum_{n=-\infty}^{\infty}\delta(t-\text{ESP}-2n\text{ESP})\right\}\right] *$$
$$\mathcal{F}\{\Pi(t/T_{\text{ETL}})\}$$
$$= \left\{ Ge^{-i\omega\tau_{\text{ramp}}/2} \operatorname{sinc}\left(\frac{1}{2}\omega\tau_{\text{ramp}}\right) \times \right.$$
$$(\tau_{\text{flattop}}+\tau_{\text{ramp}})e^{-i\omega(\tau_{\text{flattop}}+\tau_{\text{ramp}})/2} \operatorname{sinc}\left(\frac{1}{2}\omega(\tau_{\text{flattop}}+\tau_{\text{ramp}})\right) \times$$
$$\left. \frac{2\pi}{2\text{ESP}} \sum_{n=-\infty}^{\infty}\left[\delta(\omega-\frac{2\pi}{2\text{ESP}}n) - e^{-i\omega\text{ESP}}\delta(\omega-\frac{2\pi}{2\text{ESP}}n)\right] \right\} *$$
$$T_{\text{ETL}}e^{-i\omega T_{\text{ETL}}/2} \operatorname{sinc}\left(\frac{1}{2}\omega T_{\text{ETL}}\right).$$

Note that the two infinite sums on the line before last are the same apart for a factor of $e^{-i\omega\text{ESP}}$, so we can replace them by a single infinite sum multiplied by

$$1 - e^{-i\omega\text{ESP}}. \tag{S29}$$

However, the infinite sum is zero, unless $\omega = n\pi/\text{ESP} = n \cdot \omega_{2\text{ESP}}$, in which case the last expression (S29) is either zero when $n$ is even, or two, if $n$ is odd. Thus, after performing the convolution, we can write

$$\mathcal{F}\left\{G_{\text{trap.}}^{(\text{ETL})}(t)\right\} = \frac{2\pi}{\text{ESP}} G \sum_{n=-\infty}^{\infty} \left[ e^{-i(2n+1)\omega_{2\text{ESP}}\tau_{\text{ramp}}/2} \operatorname{sinc}\left(\frac{1}{2}(2n+1)\omega_{2\text{ESP}}\tau_{\text{ramp}}\right) \times \right. \quad (\text{S30})$$

$$(\tau_{\text{flattop}} + \tau_{\text{ramp}}) e^{-i(2n+1)\omega_{2\text{ESP}}(\tau_{\text{flattop}}+\tau_{\text{ramp}})/2} \times$$

$$\operatorname{sinc}\left(\frac{1}{2}(2n+1)\omega_{2\text{ESP}}(\tau_{\text{flattop}} + \tau_{\text{ramp}})\right) \times$$

$$\left. T_{\text{ETL}} e^{-i[\omega - (2n+1)\omega_{2\text{ESP}}]T_{\text{ETL}}/2} \operatorname{sinc}\left(\frac{1}{2}[\omega - (2n+1)\omega_{2\text{ESP}}]T_{\text{ETL}}\right) \right].$$

This result is an infinite sum of weighted sincs (only the sinc on the last line is a function $\omega$), centered about odd multiples of $\omega_{2\text{ESP}}$. Thus, it is the sum of all odd harmonics.

Taking the absolute value of (S27) and (S30), and *assuming* the different harmonics are sufficiently far apart that they hardly affect each other, we can approximate the absolute values as

$$\left|\mathcal{F}\left\{G_{\text{sine}}^{(\text{ETL})}(t)\right\}\right| \approx |\pi G T_{\text{ETL}}| \sum_{n=-1,0} \left|\operatorname{sinc}\left(\frac{1}{2}[\omega - (2n+1)\omega_{2\text{ESP}}]T_{\text{ETL}}\right)\right| \quad (\text{S31})$$

and

$$\left|\mathcal{F}\left\{G_{\text{trap.}}^{(\text{ETL})}(t)\right\}\right| \approx \left|\frac{2\pi}{\text{ESP}} G(\tau_{\text{flattop}} + \tau_{\text{ramp}})T_{\text{ETL}}\right| \times \quad (\text{S32})$$

$$\sum_{n=-\infty}^{\infty} \left|\operatorname{sinc}\left(\frac{1}{2}(2n+1)\omega_{2\text{ESP}} \cdot \tau_{\text{ramp}}\right) \operatorname{sinc}\left(\frac{1}{2}(2n+1)\omega_{2\text{ESP}}(\tau_{\text{flattop}} + \tau_{\text{ramp}})\right)\right| \times$$

$$\left|\operatorname{sinc}\left(\frac{1}{2}[\omega - (2n+1)\omega_{2\text{ESP}}]T_{\text{ETL}}\right)\right|$$

or

$$\left|\mathcal{F}\left\{G_{\text{sine}}^{(\text{ETL})}(t)\right\}\right| \approx \sum_{n=-\infty}^{\infty} |A_{\text{sine},n}(\omega_{2\text{ESP}})| \left|\operatorname{sinc}\left(\frac{1}{2}[\omega - (2n+1)\omega_{2\text{ESP}}]T_{\text{ETL}}\right)\right|$$

and

$$\left|\mathcal{F}\left\{G_{\text{trap.}}^{(\text{ETL})}(t)\right\}\right| \approx \sum_{n=-\infty}^{\infty} |A_{\text{trap.},n}(\omega_{2\text{ESP}})| \left|\operatorname{sinc}\left(\frac{1}{2}[\omega - (2n+1)\omega_{2\text{ESP}}]T_{\text{ETL}}\right)\right|,$$

where

$$|A_{\text{sine},n}(\omega_{2\text{ESP}})| = \begin{cases} |\pi G T_{\text{ETL}}| & n = -1, 0 \\ 0 & \text{otherwise} \end{cases} \quad (\text{S33})$$

and

$$|A_{\text{trap.},n}(\omega_{2\text{ESP}})| = \left|2\pi G \frac{\tau_{\text{flattop}} + \tau_{\text{ramp}}}{\text{ESP}} T_{\text{ETL}}\right| \times \quad (\text{S34})$$

$$\left|\operatorname{sinc}\left(\frac{1}{2}(2n+1)\omega_{2\text{ESP}} \cdot \tau_{\text{ramp}}\right) \operatorname{sinc}\left(\frac{1}{2}(2n+1)\omega_{2\text{ESP}}(\tau_{\text{flattop}} + \tau_{\text{ramp}})\right)\right|.$$

With these definitions, the absolute value of (S25) gives the simplified form

$$|g(\omega)| \approx \sum_{n=-\infty}^{\infty} |A_n(\omega_{2\text{ESP}})| \left|\operatorname{sinc}\left(\frac{1}{2}[\omega - (2n+1)\omega_{2\text{ESP}}]T_{\text{ETL}}\right)\right| \times \quad (\text{S35})$$

$$\left|\frac{\sin\left((\omega - \omega_\theta)\Delta TE \cdot N_{\text{TE}}/2\right)}{\sin\left((\omega - \omega_\theta)\Delta TE/2\right)}\right| \times \left|\frac{\sin\left(\omega \Delta T_{\text{slice}} \cdot N_{\text{slice}}/2\right)}{\sin\left(\omega \Delta T_{\text{slice}}/2\right)}\right| \times \left|\frac{\sin\left(\omega TR \cdot N_{\text{TR}}/2\right)}{\sin\left(\omega TR/2\right)}\right|,$$

where $A_n$ is either $|A_{\text{sine},n}|$ from (S33) or $|A_{\text{trap.},n}|$ from (S34), and $\omega_\theta$ is zero if all echoes are the identical, or $\pi/\Delta TE$ if the gradients switch sign every echo (see Eqs. (S9) and (S22)). This matches the expressions in the main text when $\omega_\theta = 0$, i.e., all echoes are the same.